\documentclass[11pt,twoside]{article}
\usepackage{latexsym,amssymb,amsmath,amsthm}
\usepackage{a4wide}
\usepackage[colorlinks=true]{hyperref}
\usepackage[svgnames]{xcolor}
\usepackage{tikz,tikz-cd}
\usepackage{titlesec} 
\usepackage{titling} 
\usepackage{times}
\usepackage{fancyhdr}
\usepackage{srcltx}
\usepackage{caption}
 \usepackage{subfigure}
 \usepackage{graphicx}

\DeclareMathAlphabet{\mathcal}{OMS}{cmsy}{m}{n}
\hypersetup{citecolor=DarkSlateBlue,linkcolor=DarkSlateBlue} 

\def\titlecolour{\color{DarkSlateBlue}}
\def\headcolour{\color{DarkSlateBlue}} 
\def\seccolour{\color{DarkSlateBlue}}
\def\thmcolour{\color{SlateBlue}}

\pretitle{\begin{center}\LARGE\bfseries\sffamily\titlecolour}
\preauthor{\begin{center}\large\sffamily\titlecolour}
\postauthor{\par\end{center}}

\titleformat{\section}{\Large\bfseries\sffamily\seccolour}{\thesection}{1em}{}
\titleformat{\subsection}{\large\bfseries\sffamily\seccolour}{\thesubsection}{1em}{}
\titleformat{\subsubsection}{\bfseries\sffamily\seccolour}{\thesubsubsection}{1em}{}
\titleformat{\paragraph}[runin]{\small\bfseries\sffamily\seccolour}{}{1em}{}[.]

\numberwithin{equation}{section}
\numberwithin{figure}{section}
\numberwithin{table}{section}

\title{Attracting and repelling 2-body problems on a family of surfaces of constant curvature}
\author{Luis C. Garc\'ia-Naranjo \& James Montaldi}

\newcounter{todo}

\makeatletter

\newcommand\listtodoname{List of todos}
\newcommand\listoftodos{%
  \section*{\listtodoname}\@starttoc{tod}}
\makeatother
\newtheoremstyle{theoremsf}
{2ex}
{2ex}
{\it}
{}
{\bfseries\itshape\thmcolour}
{.}
{1em}
{}

\newtheoremstyle{definitionsf}
{2ex}
{2ex}
{}
{}
{\bfseries\itshape\thmcolour}
{.}
{1em}
{}

\theoremstyle{theoremsf}
\newtheorem{theorem}{Theorem}[section]
\newtheorem{lemma}[theorem]{Lemma}
\newtheorem{proposition}[theorem]{Proposition}

\theoremstyle{definitionsf}

 \newtheorem{remark}[theorem]{Remark}

\renewenvironment{proof}{\addvspace\baselineskip\noindent{\it\thmcolour
    Proof:}\ }{\hspace*{\fill} $\Box$\par\addvspace\baselineskip}

\def\defn#1{{\bfseries\itshape #1}}

\def\d{\mathrm{d}}
\def\rr{\mathbf{r}}
\def\uu{\mathbf{u}}
\def\vv{\mathbf{v}}
\def\xx{\mathbf{x}}
\def\XX{\mathbf{X}}
\def\m{\mathbf{m}}

\def\R{\mathbb{R}}

\def\GL{\mathbf{GL}}
\def\OO{\mathbf{O}}
\def\SO{\mathbf{SO}}
\def\SE{\mathbf{SE}}

\def\re{\textsc{re}}
\def\RE {\textsc{re}}

\def\gg{\mathfrak{g}}
\def\se{\mathfrak{se}}
\def\so{\mathfrak{so}}
\def\ssl{\mathfrak{sl}}

\def\A{\mathcal{A}}
\def\SSk{\mathcal{S}_\kappa} 
\def\Lag{\mathcal{L}}
\def\Mass{\mathbb{M}}

\def\Ck{\cos_\kappa}
\def\Sk{\sin_\kappa}

\begin{document}
\maketitle
\thispagestyle{empty}

{\small
{\titlecolour\hrule height 1pt}

\medskip

\noindent{\large\sffamily\titlecolour Abstract}

\medskip

\noindent 
We first provide a classification of the pure rotational motion of 2 particles on a sphere interacting via a repelling potential.  This is achieved by providing a simple geometric equivalence between repelling particles and attracting particles, and relying on previous work on the similar classification for attracting particles. The second theme of the paper is to study the 2-body problem on a surface of constant curvature treating the curvature as a parameter, and with particular interest in how families of relative equilibria and their stability behave as the curvature passes through zero and changes sign. We consider two cases: firstly one where the particles are always attracting throughout the family, and secondly where they are attracting for negative curvature and repelling for positive curvature, interpolated by no interaction when the curvature vanishes.  
Our analysis clarifies the role of curvature in the existence and stability of relative equilibria.
\medskip


\medskip

{\titlecolour\hrule height 1pt}
}

\tableofcontents

\newpage

\section{Introduction}
\label{sec:introduction} 

This paper is concerned with the dynamics of particles on surfaces of constant Gaussian curvature,
a topic that has received much attention recently (see \cite{BMK04,D,DPR,BMB, GMPR, BGMM, DSZ,Arathoon} among others). We refer the reader to \cite{D, BMB} for
 a general introduction, a more complete set of references and  historical details of the problem.

 Our paper has two main themes. The first is a short study of the behaviour of two particles moving on a sphere which repel each other, and in particular of their purely rotational motions, complementing earlier studies of pairs of attracting particles. The second theme is fitting this into families of solutions
 that persist as the curvature varies, and particularly the study of how the dynamics or stability varies as the curvature passes through zero.  

For the first theme, if on a sphere, a particle is repelled by another particle, then it is necessarily attracted to the point antipodal to the second particle;  we formalize this in Lemma \ref{lemma:anitpodal} in order to deduce the dynamics of the repelling particles from previous studies of the dynamics of attracting particles
 (although it is evidently also possible to prove this by direct calculations, as we do Section\,\ref{sec:RE}).

One distinction between the attracting and repelling cases is that in the attracting case the particles rotate in the same hemisphere, while in the repelling case they rotate in opposite hemispheres (i.e., on opposite sides of the equator relative to the axis of rotation). See Figures\,\ref{fig:isosceles and right-angled RE} and \ref{fig:acute and obtuse RE} below. 

The principal results of this part are that if the masses are distinct, then there are two families of relative equilibria (\re), called obtuse and acute according to the angle of separation.  On the other hand, if the masses are identical, there are also two families, one, called isosceles, where the two particles subtend the same angle with the axis of rotation (though in opposite hemispheres), and the other, the `right-angled' relative equilibria, where the angle subtended at the centre of the sphere by the particles is a right angle.  This classification, and the stability of the different relative equilibria, are deduced from the corresponding results for attracting potentials \cite{BGMM}. 

It is not hard to see (and we prove below) that if the  curvature is non-positive  there are no relative equilibria for a repelling potential.

In the second part, from Section\,\ref{sec:family} on, we consider the dynamics (in particular, the relative equilibria  and their linear stability) for all surfaces of constant curvature; both attracting and repelling, and in the case of the plane, also with no interaction.
 We consider these as a function of the Gauss curvature $\kappa$, and the aim is to understand how these continue from $\kappa<0$ through $\kappa=0$ to $\kappa>0$.    
 
Our work assumes that the interaction potential between the particles $V_\kappa(q)$ depends smoothly on the curvature $\kappa$
and the distance $q$ between the particles. We also assume that for a fixed value of $\kappa$, the 
sign of   $V'_\kappa(q)$ is constant as a function of $q$. Under this  assumption 
the geometry of all of the \re\ of the problem is independent of the specific form of 
the potential whose only influence is to determine the speed of the motion.
 On the other hand, the  stability of each \re\ is sensitive to the form of the potential  
and for the stability  we 
   restrict our attention to potentials  $V_\kappa(q)$ defined via the function
 $\cot_\kappa(q)$, that analytically  interpolates between $\cot(q)$ (for $\kappa=1$) and $\coth(q)$ (for $\kappa=-1$), passing through $\cot_0(q)=q^{-1}$\,---see Section\,\ref{sec:Eq of motion} for the precise definition.

While the properties of relative equilibria  (\re) for each value of $\kappa$ are known, the question of how the families continue as $\kappa$ is varied from negative through zero to positive values has not  hitherto been considered.  The immediate difficulty that arises is that the well-known reduction to the centre of mass frame, that  reduces the 2-body problem to a central force problem when $\kappa=0$, is not available for general $\kappa$: the motion of any candidate  centre of mass does not decouple from the internal motion of the 2 particles.  In other words, Galilean relativity is not applicable to situations with non-zero curvature.

Our approach therefore is to consider only the translation and rotation symmetries: for $\kappa\neq0$ the symmetry groups are $\SO(3)$ (for $\kappa>0$) and $\SO(2,1)$ for $\kappa<0$.  For the flat case $\kappa=0$, the symmetry group is the Euclidean group $\SE(2)$.
Following previous work \cite{MoTo}, we show how these can be continued in a smooth fashion as a function of the curvature $\kappa$, together with their action on the surfaces of constant curvature $\kappa$.  
Inspired by \cite{BGMM}, we factor out this symmetry
 and obtain the reduced equations of motion, on a 5-dimensional Poisson phase space  that varies smoothly with
  $\kappa$, and where the distance $q$ between the particles is a
  phase variable. The  \re\  correspond  to equilibrium points of these reduced equations and our setup allows us to analyse their  properties as  $\kappa$ varies through zero.

 We consider two distinct families of \re\ parametrized by $\kappa$ described below, and which are 
respectively 
treated in sections~\ref{S:attracting} and~\ref{S:attracting-repelling}. These families are parametrized by 
the constant distance $q$
between the particles along the \re, and we study the behaviour of each of these  \re\ as the curvature $\kappa$ is varied through 0.

\begin{description}

\item[Attracting family:] For this family  the interaction potential $V_\kappa$  is attracting for all $\kappa$. 

For zero curvature, the   \re\ under consideration
are the  well-known uniform circular motions of the classical 2-body problem, that  we refer
 to as `Keplerian' \re. As is known, they exist for any distance $q>0$ and
 they arise due to a balance of  centrifugal and attracting forces. These  \re\
are stable in the sense that a small change in initial condition leads to small changes in the distance
between the particles throughout the motion.

Our analysis clarifies how, for arbitrary $q>0$,  these Keplerian \re\   continue smoothly  for non-zero values of the curvature. For $\kappa<0$ these are the
so-called  `elliptic' \RE\  previously found in \cite{DPR, GMPR, BGMM}, while for  $\kappa>0$ these correspond to  the
\re\   termed  `acute' (or `isosceles acute' if the masses are equal) in \cite{BGMM}, and whose existence was first indicated
in \cite{BMK04}.

For  the specific potential 
  $V_\kappa(q)=-\cot_\kappa(q)$,
we  perform the linear stability analysis  
about the corresponding equilibrium points on the 5-dimensional reduced space.  At $\kappa=0$,  there is one zero eigenvalue (with eigenvector tangent to the curve of \re), and 4 double imaginary eigenvalues  $\pm i\omega$, where $2\pi/\omega$ is the period of the circular \re\ in question. As $\kappa$ is increased or decreased from 0, the 0 eigenvalue remains, but the double eigenvalues `detune', remaining on the imaginary axis (at least for sufficiently small values of $|\kappa|$ for a fixed distance $q$ between the particles); this is illustrated in Figure\,\ref{fig:attracting-attracting}. In particular all of these \re\ are linearly stable.
  
The results described above agree with our intuition that would 
suggest that  
the Keplerian \re\ are robust under a small curvature perturbation of the ambient space as long as the distance  
between the particles is not too large. Our analysis shows that the relevant, non-dimensional, quantity that should be small is $q\sqrt{|\kappa|}$. 

Interestingly, the effects of the curvature become relevant when one attempts 
to establish the nonlinear stability of these 
\re. 
On the one hand, for $\kappa <0$ and small values of $q\sqrt{|\kappa|}$, the corresponding equilibria on the reduced
space are Lyapunov stable
on each symplectic leaf~\cite{GMPR,BGMM}.\footnote{The analysis in ~\cite{GMPR,BGMM} is for $\kappa =\pm 1$ but
may be extended to all $\kappa\neq 0$ by Remark~\ref{rmk:rescaling curvature} in our text.}
On the other hand,  for  $\kappa>0$, and even if $q\sqrt{|\kappa|}$ is small, these are  of mixed symplectic, 
or Krein sign~\cite{BGMM} (the Hessian of the reduced Hamiltonian  at a fixed symplectic leaf
 is positive definite for $\kappa<0$ but not for $\kappa>0$).\footnote{This phenomenon
 is possible since the phase space is a Poisson manifold, so standard bifurcation theory for Hamiltonian systems does not apply. }

\item[Attracting/repelling:] For this family  the  interaction potential  $V_\kappa$
 is repelling for $\kappa>0$, absent for $\kappa=0$ and attracting for $\kappa<0$. 

For $\kappa =0$ there is no interaction between the particles and the  \RE\  within our family correspond  
to  arbitrary initial configurations of 2 distinct points, that are set  in motion with equal velocities in the direction perpendicular to the line joining the particles. We refer to these as  `perpendicular  \re'. 
 Our analysis shows that if the magnitude of the velocity is tuned
appropriately with respect to  the rate at which the potential force  $F_\kappa=-V_\kappa'(q)$ vanishes at $\kappa=0$, then these \re\ persist as $\kappa$ varies from $0$ for any value of the separation between the particles. 
The underlying  reason\footnote{For negative curvature, this 
explanation appears already in \cite{GMPR}.} is that the repelling (respectively
attractive) potential forces balance the tendency of the particles to focus (respectively defocus) when the
curvature is positive (respectively negative).  As a result,
the particles travel maintaining a constant distance along curves that at every time 
are perpendicular to the geodesic
that joins them. This mechanism is illustrated in 
  Figure~\ref{fig:attractive-repelling-sketch}.

\begin{figure}[h]
\centering
\includegraphics[height=3.3cm]{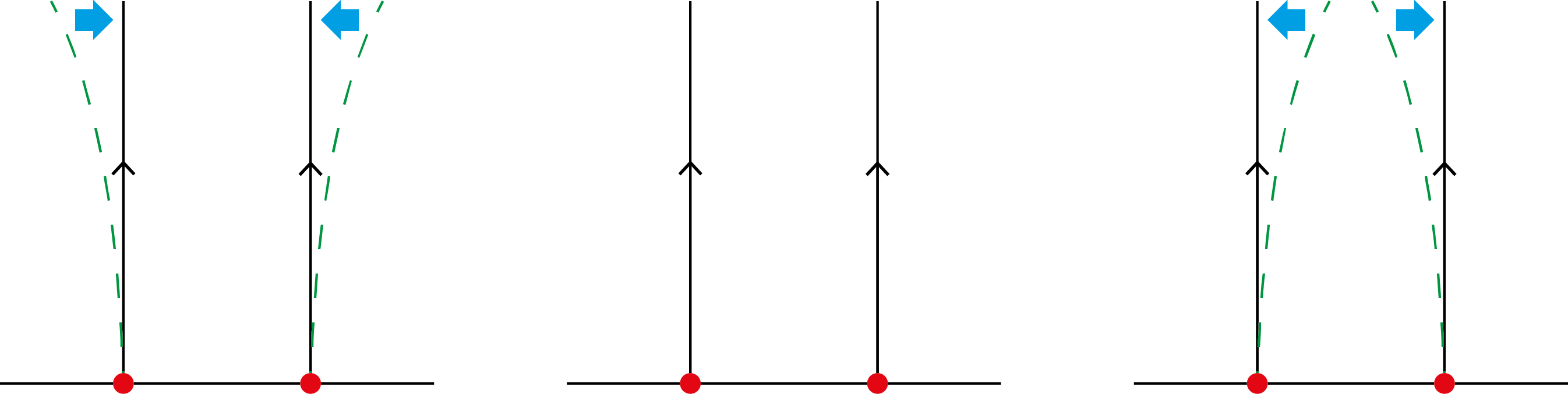}  
\put(-355,-12){$\kappa<0$ (attractive) }
\put(-233,-12){$\kappa=0$ (no interaction) }
\put(-90,-12){$\kappa>0$ (repelling) }
\caption{Sketch  of \re\ in the
attracting-repelling family. The potential force, indicated with thick arrows,
balances the tendency of the particles to defocus/focus for negative/positive curvature. } \label{fig:attractive-repelling-sketch}
\end{figure}

For $\kappa <0$ these motions correspond to the `hyperbolic' \re\ considered before in
 \cite{DPR, GMPR, BGMM}. On the other hand, for $\kappa >0$ the motion corresponds to the `acute'
 repelling \re\ whose existence is noticed  in the first part of this paper (section \ref{sec:repelling RE}).

Our analysis of this family, together with our observation that the repelling and attractive 2-body problems
on the sphere are equivalent  (Lemma \ref{lemma:anitpodal}), provides an explanation of the
mechanism responsible for the existence of obtuse  \re, studied in \cite{BMK04, BGMM}, for the attractive
2-body problem on the sphere.  

The balance of the potential forces with curvature  described above, and illustrated in
Figure~\ref{fig:attractive-repelling-sketch}, is very delicate and
as a result these \re\ are unstable for any separation $q$ when $\kappa \leq 0$ 
and for small values of $q\sqrt{\kappa}$
if $\kappa>0$. For the specific potential  $V_\kappa(q) =\kappa \cot_\kappa(q)$, the  linear approximation 
 of the 5-dimensional reduced dynamics at $\kappa=0$ turns out to be nilpotent,
  of rank 2 and satisfying $L^2=0$.  As 
 $\kappa$ varies away from $0$, one zero eigenvalue remains, while the other 4 split into a pair of real
 and a pair of purely imaginary eigenvalues (see Figure~\ref{fig:attracting-repelling}).
\end{description}

We finish the introduction by mentioning  that the treatment of the  $N$-body problem on spaces of constant curvature, with the curvature $\kappa$ as parameter was  recently considered by 
 Diacu and collaborators  \cite{D2, DIS,D3}. The emphasis of the works \cite{DIS,D3} is on deriving 
 a set of equations of motion for the problem that depend  smoothly on   $\kappa$.   However, their approach seems unnecessarily complicated, and the results become straightforward using the approach we adopt in Sec.\,\ref{sec:family} below. For example, it becomes self-evident that the equations of motion continue from $\kappa>0$ through $\kappa=0$ to $\kappa<0$. 
Moreover, in contrast to these references, our approach is naturally adapted to analysing changes in the curvature while keeping the distance between the masses fixed, which is essential  to understanding the 
 behaviour of the solutions as a function of $\kappa$.

\section{Relative equilibria for repelling particles}
\label{sec:repelling RE} 
Consider two particles, of masses $\mu_1,\mu_2$, on the unit sphere $S$ in $\R^3$ interacting via a potential energy $V(q)$, where $q$ is the geodesic distance between the particles. We assume the interaction is repelling, which is equivalent to $V'(q)<0$. 

The configuration space is $Q=S\times S \subset\R^3\times\R^3$. Given $(\xx_1,\xx_2)\in Q$, the distance $q$ between them is the unique value $q\in[0,\pi]$ satisfying $\cos q=\xx_1\cdot\xx_2$. 

\begin{figure}[t] 
\centering
\subfigure[Attracting potential]{
	\begin{tikzpicture}[scale=1.5]
		\draw [thick] (0,0) circle (1cm);
			\draw [-latex,dashed] (0,-1) -- (0,1.4); 
			\draw [-latex] (0.2,1.15) arc (-30:250: 2mm and 0.5mm); 
			\draw (-0.3,1.15) node {$\omega$};
		\draw [thin] (0,0) -- (.8660, .5);
		\draw (0.1732,0.1) arc (30:90:2mm);
	    	\filldraw [red] (.8660, .5) circle (2pt);
		\draw [thin] (0,0) -- (-.8660, .5);
		\draw (0,0.2) arc (90:150:2mm);
		\filldraw [red] (-.8660, .5) circle (2pt);
 \end{tikzpicture}
 	\hskip 5mm
\begin{tikzpicture}[scale=1.5]
		\draw [thick] (0,0) circle (1cm);
			\draw [-latex,dashed] (0,-1) -- (0,1.4); 
			\draw [-latex] (0.2,1.15) arc (-30:250: 2mm and 0.5mm); 
			\draw (-0.3,1.15) node {$\omega$};
		\draw [thin] (0,0) -- (.3418, .9398);
    		\filldraw [red] (.3418, .9398) circle (2pt);
		\draw [thin] (0,0) -- (-.9397, .3421);
		\filldraw [red] (-.9397, .3421) circle (2pt);
		\draw  (0.0683, .18796) -- (-.11958, .25638) -- (-.18794, 0.06842);
	\draw (0.1,0.45) node {$\theta$};
\end{tikzpicture}
}
 	\hskip 15mm
\subfigure[Repelling potential]{
\begin{tikzpicture}[scale=1.5]
		\draw [thick] (0,0) circle (1cm);
			\draw [-latex,dashed] (0,-1) -- (0,1.4); 
			\draw [-latex] (0.2,1.15) arc (-30:250: 2mm and 0.5mm); 
			\draw (-0.3,1.15) node {$\omega$};
		\draw [thin] (0,0) -- (.8660, .5);
		\draw (0.1732,0.1) arc (30:90:2mm);
		\filldraw [red] (.8660, .5) circle (2pt);
		\draw [thin] (0,0) -- (.8660, -.5);
		\draw (0.1732,-0.1) arc (-30:-90:2mm);
		\filldraw [red] (.8660, -.5) circle (2pt);
 \end{tikzpicture}
 	\hskip 5mm
\begin{tikzpicture}[scale=1.5]
		\draw [thick] (0,0) circle (1cm);
			\draw [-latex,dashed] (0,-1) -- (0,1.4); 
			\draw [-latex] (0.2,1.15) arc (-30:250: 2mm and 0.5mm); 
			\draw (-0.3,1.15) node {$\omega$};
		\draw [thin] (0,0) -- (.3418, .9398);
		\filldraw [red] (.3418, .9398) circle (2pt);
		\draw [thin] (0,0) -- (.9397, -.3421);
		\filldraw [red] (.9397, -.3421) circle (2pt);
		\draw  (0.0683, .18796) -- (.25630, .11954) -- (.18794, -0.06842);
	\draw (0.1,0.45) node {$\theta$};
 \end{tikzpicture}
}
 \caption{Isosceles and right-angled \re\ for identical masses for the two types of potential---all rotating about the vertical axis}
 \label{fig:isosceles and right-angled RE}
\end{figure}
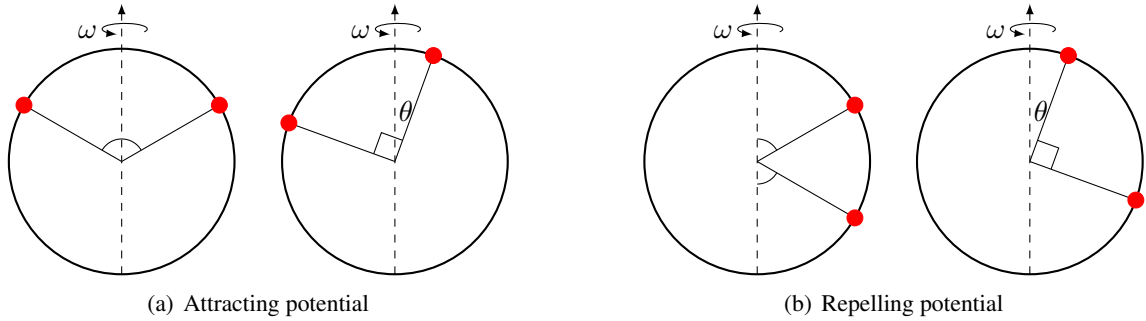

\begin{figure}[t] 
\centering
\subfigure[Attracting potential]{
	\begin{tikzpicture}[scale=1.5]  
		\draw [thick] (0,0) circle (1cm);
			\draw [-latex,dashed] (0,-1) -- (0,1.4); 
			\draw [-latex] (0.2,1.15) arc (-30:250: 2mm and 0.5mm); 
			\draw (-0.3,1.15) node {$\omega$};
			\draw [thin] (0,0) -- (.6012, .7991);
			\draw (0,0.2) arc (90:53.1:2mm);
			\draw (0.9, .8) node {$\mu_1$};
	    	\filldraw [red] (.6012, .7991) circle (1.5pt);
			\draw [thin] (0,0) -- (-.3915, .9202);
			\draw (0,0.2) arc (90:113:2mm);
			\draw (-0.7, .94) node {$\mu_2$};
	    	\filldraw [red] (-.3915, .9202) circle (2.5pt);
	 \end{tikzpicture}
	 	\hskip 3mm
	\begin{tikzpicture}[scale=1.5] 
		\draw [thick] (0,0) circle (1cm);
			\draw [-latex,dashed] (0,-1) -- (0,1.4); 
			\draw [-latex] (0.2,1.15) arc (-30:250: 2mm and 0.5mm); 
			\draw (-0.3,1.15) node {$\omega$};
			\draw [thin] (0,0) -- (.7993, .6010);
			\draw (0,0.2) arc (90:37:2mm);
			\draw (1., .65) node {$\mu_1$};
			\draw (0.2,0.35) node {$\theta_1$};
    		\filldraw [red] (.7993, .6010) circle (1.5pt);
			\draw [thin] (0,0) -- (-.9204, .3911);
			\draw (0,0.2) arc (90:157:2mm);
			\draw (-1, .64) node {$\mu_2$};
			\draw (-0.2,0.35) node {$\theta_2$};
    		\filldraw [red] (-.9204, .3911) circle (2.5pt);
 \end{tikzpicture}
 }
 	\hskip 8mm
\subfigure[Repelling potential]{
	\begin{tikzpicture}[scale=1.5]  
		\draw [thick] (0,0) circle (1cm);
			\draw [-latex,dashed] (0,-1) -- (0,1.4); 
			\draw [-latex] (0.2,1.15) arc (-30:250: 2mm and 0.5mm); 
			\draw (-0.3,1.15) node {$\omega$};
			\draw [thin] (0,0) -- (.6012, .7991);
			\draw (0,0.2) arc (90:53.1:2mm);
			\draw (0.85, .85) node {$\mu_1$};
	    	\filldraw [red] (.6012, .7991) circle (1.5pt);
			\draw [thin] (0,0) -- (.3915, -.9202);
			\draw (0,-0.2) arc (-90:-67:2mm);
			\draw (0.7, -.94) node {$\mu_2$};
	    	\filldraw [red] (.3915, -.9202) circle (2.5pt);
	 \end{tikzpicture}
	 	\hskip 3mm
	\begin{tikzpicture}[scale=1.5] 
		\draw [thick] (0,0) circle (1cm);
			\draw [-latex,dashed] (0,-1) -- (0,1.4); 
			\draw [-latex] (0.2,1.15) arc (-30:250: 2mm and 0.5mm); 
			\draw (-0.3,1.15) node {$\omega$};
			\draw [thin] (0,0) -- (.7993, .6010);
			\draw (0,0.2) arc (90:37:2mm);
			\draw (1., .65) node {$\mu_1$};
			\draw (0.2,0.35) node {$\theta_1$};
    		\filldraw [red] (.7993, .6010) circle (1.5pt);
			\draw [thin] (0,0) -- (.9204, -.3911);
			\draw (0,-0.2) arc (-90:-23:2mm);
			\draw (1, -.64) node {$\mu_2$};
			\draw (0, -1) node {};
			\draw (0.2,-0.35) node {$\theta_2$};
    		\filldraw [red] (.9204, -.3911) circle (2.5pt);
 \end{tikzpicture}
 }
 \caption{Acute and obtuse \re\ for distinct masses for the two types of potential---both rotating about the vertical axis. The configurations shown are solutions for the mass ratio $\mu_1=0.75\,\mu_2$; they have $q=\pi/3$ or $2\pi/3$ for the acute and obtuse configurations respectively. The configurations on the right are obtained from those on the left by applying the antipodal map $\A$.}
 \label{fig:acute and obtuse RE}
\end{figure}
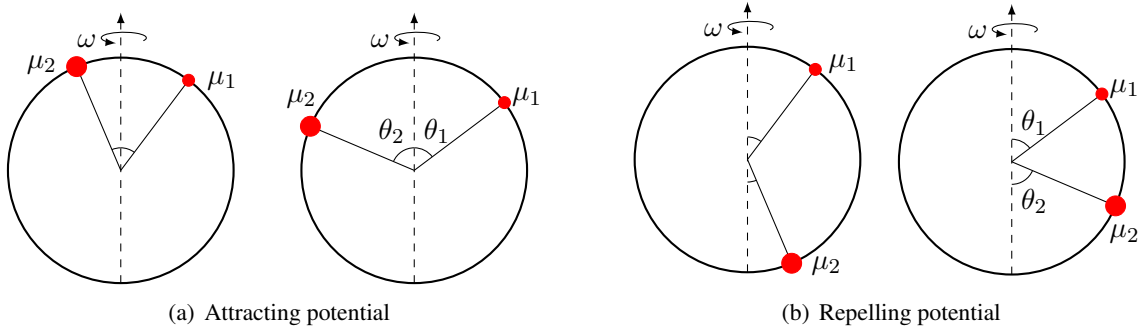

Since the potential depends only on the distance $q$, the system is symmetric under the group of rigid rotations of the sphere, that is, under $\SO(3)$. Relative equilibria are those motions corresponding to 1-parameter subgroups of $\SO(3)$, which are all rotations at some speed around a fixed axis. We now state two theorems on the classification and stability of the relative equilibria for the repelling 2-body problem. The first is for the system with distinct masses, while the second is the analogous statement for equal masses. In every case, the two masses lie on the same side of the axis of rotation, as shown in Figures\,\ref{fig:isosceles and right-angled RE} and \ref{fig:acute and obtuse RE}.

\begin{theorem}[Equal masses]
\label{Th:EqualMasses}
If the two particles are of equal mass, there are two classes of \RE, isosceles and right-angled (see Fig.\,\ref{fig:isosceles and right-angled RE}) as follows:

\begin{enumerate}
\item Given any $q\in(0,\pi)$, $q\neq\pi/2$, there is a unique \RE\ where the masses are separated by an angle $q$. In this case the axis of rotation is perpendicular to the 
sphere radius that passes midway between the masses; these we call  \emph{isosceles \RE}. 

\item Given any $\theta\in(0,\pi/4 ]$ there is a unique \RE\ with angular separation $q=\pi/2$, called a \emph{right-angled \RE}, where $\theta$ is the smaller of the angles between the axis of rotation and the masses. 
\end{enumerate}
Note that when $q=\pi/2$ and $\theta=\pi/4$ these two families meet in a pitchfork bifurcation, giving just one \RE.  The angular velocities of the \RE\ are given by \eqref{eq:omegaSphere} below.

For the specific repelling potential $V(q)=\cot q$, the linear stability of the \re\ is as follows. 
The isosceles \RE\ subtending an acute angle $q\in (0,\pi/2)$ are all unstable, while those  subtending an obtuse angle $q\in (\pi/2, \pi)$ are elliptic.
All right-angled \RE\ with $\theta \neq \pi/4$ are elliptic. 

\end{theorem}

\begin{theorem}[Distinct masses]
\label{Th:DistinctMasses}
If the masses are distinct there are also two classes of \re, acute and obtuse, as follows, see Figure\,\ref{fig:acute and obtuse RE}.

For each $q\in(0,\pi)$, $q\neq\pi/2$, there is a unique \RE\ where the masses are separated by an angle $q$. The axis of rotation subtends angles $\theta_j\in(0,\pi/2)$ with the mass $\mu_j$ such that $q=\pi - (\theta_1+\theta_2)$ (see Figure\,\ref{fig:acute and obtuse RE}(b)) which are related by
\begin{equation}
\label{eq:RE-cond1}
\mu_1\sin (2\theta_1) = \mu_2\sin(2\theta_2).
\end{equation}
We call these \emph{acute} and \emph{obtuse} \re, accordingly as $q<\pi/2$ or $q>\pi/2$.  There is no \RE\ for $q=\pi/2$.  In the acute \RE, the smaller mass is closer to the axis of rotation, while in the obtuse \RE, the larger mass is closer.  See Fig.\,\ref{fig:acute and obtuse RE}(b).

The angular velocity of the \RE\ is given by \eqref{eq:omegaSphere} below.

For the repelling potential given by $V(q)=\cot q$, the linear stability of these \RE\ are as follows.  The obtuse \RE\ are all elliptic, while for the acute ones there is a critical angle $q^\dagger\in (0,\pi/2)$ (defined below),
 which depends on the mass ratio, for which acute \RE\ are linearly unstable for $0<q<q^\dagger$ and elliptic for $q^\dagger<q<\pi/2$.
\end{theorem}

In the setting of both theorems,  the speed of rotation $\omega$ for all \re\ is given by 
\begin{equation}\label{eq:omegaSphere}
\omega^2=\zeta^{-1}V'(q),
\end{equation}
where $\zeta=\frac12\mu_1\sin(2\theta_1)=\frac12\mu_2\sin(2\theta_2)$ and $q=\pi-(\theta_1+\theta_2)$.
The critical angle  $q^\dagger$ satisfies
\begin{equation}\label{eq:q-dagger}
	q^\dagger=\alpha^\dagger -\frac\pi2 - 
			\frac{1}{2}\sin^{-1} (\mu\sin2\alpha^\dagger)
\end{equation}
where $\mu=\mu_1/\mu_2$ and $\alpha^\dagger$ is the unique solution in $(\pi/2,\pi)$ of the equation
\begin{equation}\label{eq:alpha}
    \cos2\alpha = 2\sin^2\alpha\sqrt{1+\mu^2\sin^22\alpha}.
\end{equation}
We show how this follows from the results of \cite{BGMM} in the proof below.

As with the problem for attracting particles treated in~\cite{BGMM}, the Hamiltonian function cannot be used as a Lyapunov function to guarantee the nonlinear stability of the elliptic \RE\ of the problem. 
This is due to the non-definiteness of the Hamiltonian at these points.  For the particular repelling potential given by 
$V(q) = \cot(q)$, we find the following. 

\begin{proposition}  \label{prop:signature}
The restriction to the symplectic leaf of the Hessian matrix of the reduced Hamiltonian, with $V(q)=\cot(q)$, 
at the \RE\ of the problem described in the theorems above, has the following signature:
\begin{description}
\item[$\mu_1 = \mu_2$\,:]
\begin{enumerate}
\item Right-angled \RE\ which are not isosceles have signature $(++--)$.
\item Isosceles  \RE\ subtending an acute angle have signature $(+++-)$.
\item Isosceles  \RE\ subtending an obtuse angle have signature   $(++--)$.
\end{enumerate}
\item[$\mu_1\neq \mu_2$\,:]
\begin{enumerate}
\item Acute \RE\ with $0<q<q^\dagger$ have signature $(+++-)$.
\item Acute \RE\ with $q^\dagger<q<\pi/2$ have signature $(++--)$.
\item Obtuse \RE\  have signature $(++--)$.
\end{enumerate}
\end{description}
\end{proposition}

In the equal mass case, one can clearly see from the change in signature of the Hamiltonian, the pitchfork bifurcation occuring as momentum is increased. The central family is the isosceles \RE\ with the right-angled \RE\ bifurcating off that family at $\theta=\pi/4$. (If we use $q$ as a parameter, then the pitchfork bifurcation is a `vertical bifurcation' since the right-angled \RE\ all have $q=\pi/2$.)

In the case of distinct masses, the change in signature of the Hamiltonian at $q=q_*$ occurs at a saddle-node bifurcation where the angular momentum is minimal. 

The theorem is a direct consequence of the results of \cite{BGMM} via the following observation.

\begin{lemma}\label{lemma:anitpodal}
Define the diffeomorphism $\A:Q\to Q$ by $\A(\xx_1,\xx_2)=(\xx_1,-\xx_2)$; that is, $\xx_2$ is mapped to its antipodal point. This map transforms a repelling 2 body system on the sphere to an attracting one. Specifically, if the Lagrangian of the repelling system is
$$\Lag_1 = \tfrac12\mu_1\|\dot\xx_1\|^2 + \tfrac12\mu_2\|\dot\xx_2\|^2 - V(q)$$
then the transformed system has
$$\Lag_2 = \tfrac12\mu_1\|\dot\xx_1\|^2 + \tfrac12\mu_2\|\dot\xx_2\|^2 - V(\pi-q).$$
\end{lemma}

Note that if the first system is repelling, so $V'(q)>0$, then the second is attracting: $\frac{\d}{\d q}V(\pi-q)<0$. 

\begin{proof}
Let $\Lag_1$ be as in the statement of the lemma. If $(\xx_1,\xx_2)$ are separated by distance $q$ then $\A(\xx_1,\xx_2)$ are separated by a distance of $q'=\pi-q$.  
It follows that applying the map $\A$, or rather its lift $T\A$ to the tangent bundle, gives
$$\Lag\circ T\A = \tfrac12\mu_1\|\dot\xx_1\|^2 + \tfrac12\mu_2\|\dot\xx_2\|^2 - V(q').$$
Now $V(q')=V(\pi-q)$ is a smooth increasing function of $q$, meaning that the transformed Lagrangian describes an attracting system.  
\end{proof}

\begin{proof}(of both theorems and the proposition) 
Consider the repelling system on $Q$ with potential $V$, a smooth decreasing function of $q\in(0,\pi)$.  The Lagrangian is given by
$$\Lag = \tfrac12\mu_1\|\dot\xx_1\|^2 + \tfrac12\mu_2\|\dot\xx_2\|^2 - V(q).$$

By the lemma, the diffeomorphism $\A$ transforms the system with a repelling potential to one with an attracting potential. 
This new attracting system is precisely the subject of \cite{BGMM}, and the two theorems above follow from Theorems 4.1 and 4.3 of that paper, after exchanging acute with obtuse (while right-angled configurations map to right-angled configurations, preserving $\theta$).   For example, in \cite[Theorem 4.3]{BGMM} it is shown that for different masses, the acute \RE\ are linearly stable. By applying the map $\A$, one deduces that the obtuse \RE\ for the repelling problem are linearly stable.  
The proposition follows similarly from \cite[Proposition 4.4]{BGMM}, after noting that $\cot(\pi-q)=-\cot(q)$.  

The expression for the bifurcation value $q^\dagger$ requires more explanation.  In \cite[Sec.\,4]{BGMM} the angle $\alpha$ is introduced as the angle between mass 1 and the axis of rotation, measured in the direction towards mass 2.  It is found that the relation with $q$ is
\begin{equation}\label{eq:alpha-q}
    \mu_1\sin(2\alpha)=\mu_2\sin(2(q-\alpha)).
\end{equation}
This is a purely kinematic relation and is independent of any potential.  
For the specific potential $V=-\cot q$,  it is shown in \cite[Theorem 4.3]{BGMM} that the loss of stability  occurs when $\alpha$ lies in the interval $(0,\pi/2)$ and satisfies \eqref{eq:alpha};
this value is denoted $\alpha^*$.   The bifurcation then occurs when $q=q^*\in(\pi/2,\,\pi)$ satisfies \eqref{eq:alpha-q}.

\begin{figure}
\centering
	\begin{tikzpicture}[scale=1.5] 
		\draw [thick] (0,0) circle (1cm);  
    		\filldraw [red] (0,1) circle (1.5pt);
			\draw [thin] (0,0) -- (0,1);
		\draw (0,1) node [anchor=south] {$\mu_1$};
			\draw [thin] (0,0) -- (.9204, -.3911);
			\draw (.9204, -.3911) node [anchor=west] {$\mu_2$};
	    		\filldraw [red] (.9204, -.3911) circle (2pt);
			\draw [dashed] (-1.31558, -.47894) -- (1.31558, .47894); 
			\draw (0,0.2) arc (90:20:2mm);
			\draw (0.2,0.3) node {$\alpha$};
 \end{tikzpicture}\qquad\qquad
\begin{tikzpicture}[scale=1.5] %
		\draw [thick] (0,0) circle (1cm);  
    		\filldraw [red] (0,1) circle (1.5pt);
			\draw [thin] (0,0) -- (0,1);
		\draw (0,1) node [anchor=south] {$\mu_1$};
			\draw [thin] (0,0) -- (-.9204, .3911);
			\draw (-.9204, .3911) node [anchor=east] {$\mu_2$};
	    		\filldraw [red] (-.9204, .3911) circle (2pt);
			\draw [dashed] (-1.31558, -.47894) -- (1.31558, .47894); 
			\draw (0,0.2) arc (90:200:2mm);
			\draw (-0.2,0.3) node {$\alpha$};
 \end{tikzpicture}

\caption{The definition of $\alpha$ for the attracting system (left) and the resulting repelling configuration (right). The dashed line is the axis of rotation.}
\label{fig:alpha}
\end{figure}
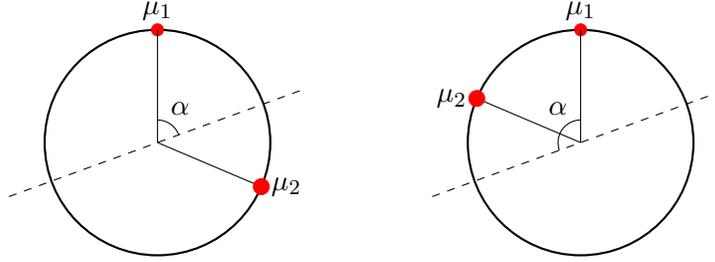

Given the same definition of $\alpha$, when applying the antipodal map $\A$, the angle $\alpha$ is changed to $\pi-\alpha$, as illustrated in Figure\,\ref{fig:alpha}.   Such $\alpha$ satisfies the same equation \eqref{eq:alpha}, but is now the solution in the interval $(\pi/2,\pi)$; call this value $\alpha^\dagger$ (equal to $\pi-\alpha^*$).  In the attracting problem, $\alpha<q$ while in the repelling problem, $\alpha>q$ (as illustrated in the figure). If $q^*>\pi/2$ is the bifurcation value determined in \cite{BGMM} (for obtuse attracting configurations), then the corresponding bifurcation for the (acute) repelling problem is $q^\dagger=\pi-q^*$. It follows that $q^\dagger$ indeed satisfies \eqref{eq:q-dagger}. 
\end{proof}

\begin{remark}
It is usual to assume that $\lim_{q\to0}V(q)$ and $\lim_{q\to \pi}V(q)$ are infinite, in which case the diagonal ($\xx_1=\xx_2$) and antipodal ($\xx_1=-\xx_2$) configurations are forbidden. In that case the \RE\ we have described above are all the possible ones.  On the other hand, without this assumption (for example $V(q)=\pm\cos q$ where `$+$' is repelling, `$-$' is attracting), the diagonal and antipodal configurations will be equilibria, and motion around a great circle at any (common) speed will be a relative equilibrium.  In the repelling case, the antipodal \RE\ and equilibria will be nonlinearly stable relative to the group of symmetries, while the diagonal ones will be unstable. For the attracting case, the existence is the same, but the stabilities are reversed. Some details in this direction can be found in \cite{BMK04}.
\end{remark}

As explained in the introduction, the mechanism responsible for the existence of the   \re\ in Theorems~\ref{Th:EqualMasses}
and~\ref{Th:DistinctMasses} that subtend an acute angle $q$ 
is the balance of the repelling forces 
with the tendency of the geodesics on the sphere to focus. The interpretation of the   \re\ that subtend an obtuse
angle is more subtle as it involves
the balance of centrifugal and attracting forces of $\mu_1$ and the `phantom attracting particle' antipodal to $\mu_2$ in
 Lemma \ref{lemma:anitpodal}.

\section{Curvature family}
\label{sec:family} 

Combining the results of Section\,\ref{sec:repelling RE} with previous work on attracting potentials on surfaces of constant curvature, it is interesting to study how the families of relative equilibria are related as the curvature $\kappa$ varies from positive, through zero to negative values. Similar studies  were done previously for the Kepler problem in \cite{CRS} and for point vortices in \cite{MoTo}, and we borrow freely from the approach taken in the latter paper. 

Denote by $\SSk$ the surface in $\R^3$ defined by 
\begin{equation} \label{eq:surface}
f_\kappa(x,y,z) = x^2+y^2+\kappa z^2-2z=0,
\end{equation}
and on this surface we consider the restriction of the ambient metric in $\R^3$ given by
\begin{equation} \label{eq:ambient metric}
\d s^2 = \d x^2+\d y^2 + \kappa\, \d z^2.	
\end{equation}
Writing $K_ \kappa = \mathrm{diag}[1,1, \kappa]$ for the metric tensor, we define the related norm by,
$$\|\xx\|_\kappa ^2 = \xx^TK_\kappa\xx,\quad\text{for } \xx\in\R^3.$$
The Riemannian surface $\SSk$ with the given metric is a surface of constant curvature $\kappa$, seen as follows. The equation \eqref{eq:surface} can also be written 
$x^2+y^2+ \kappa(z-\frac{1}{\kappa})^2=\frac1{\kappa}$.

\begin{description}
\item{$\kappa>0$:} Put $u=\sqrt{\kappa}(z-\frac1{\kappa})$; the metric becomes the usual Euclidean metric and the equation that of the sphere centred at the origin and of radius $1/\sqrt{\kappa}$, which is indeed of curvature $\kappa$. In the original coordinates, the centre is at $(x,y,z)=(0,0,1/\kappa)$. Note for future reference that for $\kappa>0$ the maximum distance between two points on $\SSk$ is $\pi/\sqrt{\kappa}$. 

\item{$\kappa=0$:} The metric on $\R^3$ is degenerate. However, restricted to the surface $\mathcal{S}_0$, which in this case is a paraboloid, it becomes a Riemannian metric, and moreover the orthogonal projection to the $x$-$y$ plane with the Euclidean metric is an isometry, showing that the metric on $\mathcal{S}_0$ has curvature zero. 
	
\item{$\kappa<0$:} Here put $u=\sqrt{-\kappa}(z-\frac1{\kappa})$, to see that the metric is the standard Minkowski metric and the surface $\SSk$ is a hyperboloid of 2 sheets. We restrict attention to the upper sheet, with $z\geq0$, and it is then standard that this surface inherits a hyperbolic metric of constant curvature $\kappa$. 
\end{description}

\begin{remark}\label{rmk:rescaling curvature}
For $\kappa\neq0$, putting $(x,y,z) = (X/\sqrt{|\kappa|},\,Y/\sqrt{|\kappa|},\,Z/\kappa)$ maps $(x,y,z)\in\SSk$ to $(X,Y,Z)\in\mathcal{S}_{\pm1}$, and changes the metric to $\frac1{|\kappa|}K_{\pm1}$.  If the potential energy scales with $\kappa$ in a power law (as is the case for the potentials we consider below, \emph{viz}.\ $V_\kappa(q)=-\cot_\kappa(q)$ and $\kappa\cot_\kappa(q)$) then, possibly after a rescaling of time depending on the power law, the Lagrangian scales and the systems for $\kappa=\kappa_0\neq0$ and $\kappa=\mathrm{sign}(\kappa_0)=\pm1$ are equivalent. 
\end{remark}

\subsection{Symmetry group} 

When $K_\kappa$ is invertible, the Lie group of linear transformations preserving the metric \eqref{eq:ambient metric} is denoted $\SO(K_\kappa)$. The elements $g$ satisfy $g^TK_\kappa g = K_\kappa$.  For the Lie algebra, we have $\xi\in\so(K_\kappa)$ if and only if $\xi^T K_\kappa + K_\kappa\xi=0$. A basis for $\so(K_\kappa)$ is given by,
\begin{equation}\label{eq:Lie algebra basis}
	\xi_1 = \begin{pmatrix}0&0&0\cr 0&0&-\kappa\cr 0&1&0\end{pmatrix},\quad
	\xi_2 = \begin{pmatrix}0&0&\kappa\cr 0&0&0\cr -1&0&0\end{pmatrix},\quad
	\xi_3 = \begin{pmatrix}0&-1&0\cr 1&0&0\cr 0&0&0\end{pmatrix}.
\end{equation}
This basis satisfies the commutation relations
$$
[\xi_1,\, \xi_2] = \kappa \xi_3,\quad [\xi_2,\,\xi_3] = \xi_1,\quad 
	[\xi_3,\, \xi_1] = \xi_2.
$$
When $\kappa=0$, one sees that $g\in \GL(3)$ satisfies $g^TK_0g=K_0$ if and only if $g$ is of the form 
\begin{equation}\label{eq:SE(2) matrix}
	g = \left(\begin{array}{c|c}
	A & \begin{matrix} 0\cr 0\end{matrix} \cr
	\hline
	\begin{matrix} a&b\end{matrix} &c
\end{array}\right)
\end{equation}
where $A\in \OO(2)$, and $a,b,c\in\R$ with $c\neq0$. This is a 4-dimensional group, whereas we need the subgroup isomorphic to $\SE(2)$ (as $\kappa=0$ corresponds to the Euclidean plane), and this is the subgroup with $c=1$.   The Lie algebra of this subgroup coincides with the algebra generated by the $\xi_j$ above with $\kappa=0$.  With this definition of $\SO(K_0)$, from now on, we abbreviate $G_\kappa = \SO(K_\kappa)$ and
$\gg_\kappa=\so(K_\kappa)$.

The Lie algebra $\gg_\kappa$ is isomorphic to
$\so(3)$ for $\kappa>0$, to $\se(2)$ for $\kappa=0$, and to $\ssl(2)\simeq\so(2,1)$ for $\kappa<0$.
Indeed, for $\kappa\neq0$, the standard commutation relations are
recovered by rescaling the basis to $\{|\kappa|^{-1/2} \xi_1,\,
|\kappa|^{-1/2} \xi_2,\, \xi_3\}$.  So
$$
G_\kappa\simeq 
	\begin{cases}
		\SO(3) & \text{if } \kappa>0\cr
		\SE(2) & \text{if } \kappa=0\cr
		\SO(2,1) & \text{if } \kappa<0.
	\end{cases}
$$

\paragraph{Affine action}
 The linear action of $G_\kappa$ on $\R^3$ preserves the metric $K_\kappa$ (by construction). 
In order to preserve the surface $\SSk$, the action needs to be modified by a translation $T_\kappa(g)$ 
depending on $g\in G_\kappa$
as follows: 
\begin{equation*}
g\cdot \xx= g\xx +T_\kappa(g), 
\end{equation*}
where
$$T_\kappa(g) = \begin{cases}
	\frac1{\kappa}(I-g)\mathbf{e}_3 & \text{if } \kappa\neq0, \\[6pt]
	\begin{pmatrix}A\mathbf{u}\cr \frac12|\mathbf{u}|^2\end{pmatrix} & \text{if } \kappa=0.
	\end{cases}
$$
Here ${\mathbf{e}}_3= (0,0,1)^T$, and, for $\kappa=0$, $g$ is as in \eqref{eq:SE(2) matrix}, with $c=1$ and we write $\mathbf{u}=(a,b)^T$.  One can show that $T_\kappa(g_\kappa)$ is continuous in $\kappa$, whenever $g_\kappa$ is a continuous family of matrices with $g_\kappa\in G_\kappa$.

The corresponding affine action of the Lie algebra $\gg_\kappa$ on $\R^3$ is given by
\begin{equation}\label{eq:infinitesimal action}
\xi\cdot\xx  = \xi\xx+\tau(\xi),
\end{equation}
where $\xi\xx$ is the linear part (matrix times vector) and
\begin{equation}\label{eq:tau}
	\tau(\omega_1\xi_1+\omega_2\xi_2+\omega_3\xi_3) = \begin{pmatrix}-\omega_2\cr \omega_1\cr0\end{pmatrix}
\end{equation}
is the translation.

\paragraph{Lie Poisson structure on $\gg_\kappa^*$}
Throughout we will write $\m=(m_1,m_2,m_3)$ for a point in $\gg_\kappa^*$ using the basis dual to the basis $\{\xi_j\}$. 
It is seen from the commutation relations that the \emph{minus} Lie Poisson structure
on $\gg_\kappa^*$ is determined by 
\begin{equation*}
\{m_1,m_2\}=-\kappa m_3, \quad \{m_2,m_3\}=- m_1, \quad  \{m_3,m_1\}=- m_2.
\end{equation*}
Hence, the bracket of functions $f,g$ on $\gg_\kappa^*$ is 
\begin{equation}\label{eq:Lie-Poisson}
	\{f,\,g\}(\m) = - \left ( K_\kappa \m  ,  \nabla f \times \nabla g \right ),
\end{equation} 
where $\times$ is the ordinary vector product in $\R^3$, and this holds also in the degenerate case $\kappa=0$. It follows immediately from the above formula that the function 
\begin{equation}\label{eq:Casimir}
	C(\m)=	m_1^2+m_2^2+\kappa m_3^2
\end{equation} 
is a Casimir function of the bracket. 

\subsection{Equations of motion} \label{sec:Eq of motion}
There exist well-known functions that interpolate between the circular trigonometric  functions and the hyperbolic ones. These are often denoted $\sin_\kappa$ and $\cos_\kappa$ (or $S_\kappa$ and $C_\kappa$) for $\kappa\in\R$ and are defined as follows,
\begin{equation} \label{eq:trig fns}
	\begin{array}{rcl}
		\sin_\kappa(x) &=& \begin{cases}
 						\frac1{\sqrt{\kappa}}\,\sin(\sqrt{\kappa}\,x) & \text{if \ } \kappa >0,\cr
 						x & \text{if \ } \kappa =0,\cr
 						\frac1{\sqrt{-\kappa}}\,\sinh(\sqrt{-\kappa}\,x) & \text{if \ } \kappa <0,\cr
					 \end{cases} \\[24pt]
		\cos_\kappa(x) &=& \begin{cases}
 						\cos(\sqrt{\kappa}\,x) & \text{if \ } \kappa>0,\cr
 						1 & \text{if \ } \kappa =0,\cr
 						\cosh(\sqrt{-\kappa}\,x) & \text{if \ } \kappa <0,\cr
					 \end{cases}
	\end{array}
\end{equation}
Note that these are analytic functions of $(\kappa,x)$; indeed the Taylor series at the orgin are
\begin{equation} \label{eq:Taylor}
	\sin_\kappa(x) = x-\tfrac1{3!} \kappa x^3+\tfrac1{5!} \kappa^2x^5+\cdots,\qquad 
	\cos_\kappa(x) = 1 - \tfrac12 \kappa x^2 +\tfrac1{4!} \kappa^2x^4 +\cdots.
\end{equation}
It is easy to check that these functions satisfy similar relations to the usual trigonometric functions, for example $\cos_\kappa(x)^2 + \kappa \sin_\kappa(x)^2=1$, and 
\begin{align*}
\sin_\kappa(x+y) &= \cos_\kappa(x) \sin_\kappa(y)+ \sin_\kappa(x) \cos_\kappa(y),  \\ \cos_\kappa(x+y) &= \cos_\kappa(x) \cos_\kappa(y) -\kappa \sin_\kappa(x) \sin_\kappa(y).
\end{align*}
Moreover, the derivatives satisfy $\sin_\kappa' = \cos_\kappa$, and $\cos_\kappa'=-\kappa \sin_\kappa$.  In our application of these functions,  $\kappa$ represents the Gauss curvature, and $x$ is replaced by $q$; note that $q$ has a unit of length, but $\sqrt\kappa$ has a unit of inverse-length, so the arguments of $\sin$, $\sinh$ etc., are dimensionless.  See \cite{CRS} for a similar use of these functions in the curved Kepler problem. One also defines other functions, such as $\tan_\kappa=\sin_\kappa/\cos_\kappa$ and $\cot_\kappa$ similarly.

Now consider 2 point masses on the surface $\SSk$ of constant curvature $\kappa$.  We assume the points are distinct, and when $\kappa>0$ that the points are not antipodal.  The configuration space of this system is therefore
$$Q_\kappa = \SSk\times \SSk \setminus \Delta,$$
where $\Delta$ consists of the points on the diagonal (for all $\kappa$) together with pairs of antipodal points if $\kappa>0$. 

The analysis that follows is very close to that in \cite{BGMM}, so we omit the details of the calculations. 
  
We parametrize points on $Q_\kappa$ as $Q_\kappa\simeq I_\kappa\times G_\kappa$, where $I_\kappa$ is the interval of allowed distances between the pair of points in a configuration. That is, 
$$I_\kappa = \begin{cases}
  (0,\infty) & \text{if } \kappa\leq0\cr
  (0,\pi/\sqrt{\kappa}) & \text{if } \kappa>0.
 \end{cases}
$$
Now, given $q\in I_\kappa$ consider the two points in $\R^3$, 
$$\xx_1= \begin{pmatrix}	0\cr0\cr0  \end{pmatrix},
\qquad \xx_2 = \xx_2(q) =
	\begin{pmatrix}
		 0 \cr 	\Sk(q) \cr \frac1{\kappa}(1-\Ck(q))
 \end{pmatrix}.
$$
For $\kappa=0$ the final component of the point $\xx_2$ is defined by continuity, giving $\frac12 q^2$, as can be seen from \eqref{eq:Taylor}. 

\begin{lemma}
	The points $\xx_1,\xx_2$ lie in $\SSk$ separated by a (geodesic) distance $q$.
\end{lemma}

\begin{proof}
It is easy to check that both points lie in $\SSk$. For $t\in [0,q]$ let
$$\gamma(t) = \begin{pmatrix}
 0 \cr 	\Sk(t) \cr \frac{1}{\kappa}(1-\Ck(t))
 \end{pmatrix}. $$
Then $\gamma(0)=\xx_1$,  $\gamma(q)=\xx_2$ and $\gamma(t)\in \SSk$.  A short calculation shows that $\|\dot\gamma(t)\|_\kappa^2=1$, whence $\gamma$ is a  curve parametrized by arc length and the distance along $\gamma$ from the origin to $\xx_2$ is therefore equal to $q$. There remains to argue that $\gamma$ is a geodesic. This follows by symmetry: the surface $\SSk$ is a surface of revolution and $\gamma$ lies on a meridian. 
\end{proof}

The parametrization (diffeomorphism) of $Q_\kappa$ by $I_\kappa\times G_\kappa$ is given by
$$(q,\,g) \longmapsto (g\cdot\xx_1,\,g\cdot \xx_2 (q))\in Q_\kappa.$$
Here $g\cdot\xx$ is the affine action of $G_\kappa$ described above. To write down the Lagrangian we pass to the tangent bundles, and we use the left trivialization for $TG_\kappa = G_\kappa\times\gg_\kappa$, $(g,\dot g)\mapsto (g,\xi)=(g,g^{-1}\dot g)$. This gives
\begin{eqnarray*}
	T\,I_\kappa\times TG_\kappa & \longrightarrow & TQ_\kappa \subset T\SSk\times T\SSk \subset\R^6\times\R^6 \cr 
	(q,\dot q, g, \xi) & \longmapsto & (g\cdot\xx_1,\,g\xi\cdot\xx_1,\, g\cdot\xx_2,\,g\xi\cdot\xx_2+ g\xx_2'\dot q),
\end{eqnarray*}
where $\xx_2' = (\frac{\d}{\d q}\xx_2)$.  The fact we are using the left-trivialization of $TG$ means that $\xi$ is to be interpreted as angular velocity \emph{in the body frame} \cite{MaRa}.

Now consider a general motion.  Associated to $(q(t),g(t))$ in $I_\kappa\times G$, write
$$\XX_1(t)=g(t) \cdot \xx_1,\quad\text{and}\quad \XX_2(t)=g(t)\cdot \xx_2(q(t)).$$
Given a potential $V_\kappa(q)$, the Lagrangian $\Lag:TQ_\kappa\to\R$ is then given in terms of this parametrization as follows,
\begin{eqnarray*}
  \Lag &=& \tfrac12\mu_1\|\dot\XX_1\|_\kappa^2 + \tfrac12\mu_2\|\dot\XX_2\|_\kappa^2 -V_\kappa(q) \\
  &=& \tfrac12 \mu_1 \|g(\xi\cdot\xx_1)\|_\kappa^2 + \tfrac12\mu_2\|g(\xi\cdot\xx_2+\xx_2'\dot q)\|_\kappa^2 -V_\kappa(q)\\
  &=& \tfrac12 \mu_1 \|\xi\cdot\xx_1\|_\kappa^2 + \tfrac12\mu_2\|\xi\cdot\xx_2+\xx_2'\dot q\|_\kappa^2  -V_\kappa(q)\\
  &=& \tfrac12 \mu_1 \|\tau(\xi)\|_\kappa^2 + \tfrac12\mu_2\|\xi\xx_2+\tau(\xi)+\xx_2'\dot q\|_\kappa^2 -V_\kappa(q).
\end{eqnarray*}
Here we have used the fact that $g$ is norm-preserving. Writing,
$$\xi = \omega_1\xi_1+\omega_2\xi_2+\omega_3\xi_3,$$
by \eqref{eq:tau}, we obtain $\|\tau(\xi)\|_\kappa^2 = (\omega_1^2+ \omega_2^2)$ and a short calculation shows that,
$$\xi\xx_2+\tau(\xi)+\xx_2'\dot q = 
	\begin{pmatrix}
		- \Sk(q)\,\omega_3 -\Ck(q)\,\omega_2 \cr 
		\Ck(q)(\omega_1+\dot q) \cr 
		\Sk(q)(\omega_1+\dot q)
	\end{pmatrix} 
$$ 
whence
$$\|\xi\xx_2+\tau(\xi)+\xx_2'\dot q\|_\kappa^2 = (\omega_1+\dot q)^2 + (\Sk(q)\omega_3+ \Ck(q)\omega_2)^2.$$ 

Combining these expressions, one obtains
$$\Lag = \tfrac12 \vv^T \Mass \vv - V_\kappa(q),$$
where $\vv=(\dot q, \omega_1,\omega_2,\omega_3)$, and the mass matrix $\Mass$ is given by
\begin{equation}\label{eq:Mass matrix}
	\Mass = \begin{pmatrix}
				\mu_2 & \mu_2 &0&0 \cr
				\mu_2& \mu_1+\mu_2&0&0\cr
				0&0& \mu_1+\mu_2C^2 & \mu_2SC \cr
				0&0& \mu_2SC & \mu_2S^2
			\end{pmatrix}.
\end{equation}
Here and in the matrix below, $S=\Sk(q)$ and $C=\Ck(q)$. 
Recall that when $\kappa=0$ then $C=1$ and $S=q$. 

 To pass to the Hamiltonian formulation, we use the Legendre transform, with $p=\partial\Lag/\partial \dot q$, and $m_j = \partial\Lag/\partial \omega_j$ ($j=1,2,3$). 
This involves the left trivialization of $T^*G_\kappa$ as $G_\kappa\times\gg_\kappa ^*$ (see for example \cite{MaRa}).  
Putting $\uu=(p,m_1,m_2,m_3)$, one finds.                                                                                                                                                                                                                                                                                                                                                                                                                                                                                                                                                                                                         
\begin{equation} \label{eq:Hamiltonian}
	H =  \tfrac12\uu^T\Mass^{-1}\uu + V_\kappa(q),
\end{equation}
with 
\begin{equation}\label{eq:inverse Mass matrix}
	\Mass^{-1} = \frac{1}{\mu_1\mu_2}
			\begin{pmatrix}
			\mu_1+\mu_2 & -\mu_2 &0&0 \cr
			-\mu_2& \mu_2 &0&0\cr
			0&0& \mu_2 & -\tfrac{\mu_2C}{S} \cr
			0&0& -\tfrac{\mu_2C}{S} & \tfrac1{S^2}(\mu_1+\mu_2C^2)
		\end{pmatrix}.
\end{equation}

\paragraph{Reduction} 
The Hamiltonian $H(q,p,g,\m)$ is independent of $g\in G_\kappa$, so induces a smooth function on the smooth Poisson reduced space $P = T^*I_\kappa\times\gg_\kappa^*$, which we also denote $H=H(q,p,\m)$. The reduced equations of motion are determined by the Poisson structure arising from the left trivialization \cite{MaRa}, 
$$\{q,p\}=1,\quad \{m_1,m_2\} = -\kappa m_3,\quad  \{m_2,m_3\} = - m_1,\quad \{m_3,m_1\} = - m_2\,;$$
brackets between other coordinates being zero. The equations of motion of the reduced system are therefore,
\begin{equation}\label{eq:equations of motion}
\dot q = H_p,\quad \dot p =- H_q,\quad \text{and}\quad \dot\m = (K_\kappa\m)\times(H_\m).
\end{equation}
Both the Hamiltonian and the Casimir $C$ given by~\eqref{eq:Casimir} are first integrals.

\subsubsection{Flat case: $\kappa=0$} \label{sec:flat case momentum}
The usual approach to the 2-body problem in the plane is to fix the centre of mass (at the origin), which implies vanishing total momentum $\mathbf{p}=0$, and to use the relative vector $\rr =\xx_2-\xx_1$ for configuration space, which then satisfies a central force law (the Kepler problem).  However, that approach does not fit into the family with non-zero curvature, since these problems have no preferred centre of mass frame.  We therefore describe briefly the relation between the usual `central force' approach and the one we need to take here.  

Let 
\begin{equation*}
\mathbf{p}_1= \mu_1 \dot \xx_1, \qquad  \mathbf{p}_2= \mu_2 \dot \xx_2,
\qquad  \mathbf{p} = \mathbf{p}_1+ \mathbf{p}_2,
\end{equation*}
be the usual momenta and $\rr  = \xx_2-\xx_1$. 
The variables $(q,p,m_1,m_2,m_3)$ then correspond to 
\begin{equation} \label{eq:relations between variables}
q=\|\rr \|,\quad p= \frac{\rr  \cdot  \mathbf{p}_2 }{q}, \quad m_1= \frac{\rr  \cdot  \mathbf{p} }{q}, \quad m_2= \frac{\rr  \times  \mathbf{p} }{q}, \quad m_3= \rr \times \mathbf{p}_2,
\end{equation}
 where  $\rr \times  \mathbf{p}$ denotes the scalar quantity   given by   the third
component of the cross product of the vectors $(\rr,0),  (\mathbf{p},0) \in \R^3$ (and similarly for $\rr \times  \mathbf{p}_2$). 
In particular, the angular momentum about the centre of mass is given by
\begin{equation} \label{eq:L=ang mmtm}
L = m_3 -\frac{\mu_2\, q}{\mu_1+\mu_2}m_2,
\end{equation}
which is a first integral of the equations of motion \eqref{eq:equations of motion} in this case where $\kappa=0$, and in this case the Casimir is $C=m_1^2+m_2^2=\|\mathbf{p}\|^2$. 

\begin{remark}\label{rmk:rotation about centre of mass}
Given that $L$ is an integral, it is natural to ask what the corresponding Hamiltonian flow (or group action) is. In fact this is simply planar rotations of the plane about the centre of mass. Its expression in terms of the variables $(q,p,m_1,m_2,m_3)$ can be determined from \eqref{eq:relations between variables}.  In particular $(m_1,m_2)$ rotates about the origin, thus preserving the Casimir $C=m_1^2+m_2^2=\|\mathbf{p}\|^2$. 
\end{remark}

\subsection{Relative equilibria} 
	\label{sec:RE}
Relative equilibria are equilibrium points of the reduced system so they correspond to solutions of  the following equations (see \eqref{eq:equations of motion}):
\begin{subequations}\label{eq:RE}
\begin{align}
\label{eq:RE-p}
\frac{\partial H}{\partial p}&=0, \\
\label{eq:RE-m}
{K_\kappa\m} \times \frac{\partial H}{\partial \m}& ={\bf 0}, \\
\label{eq:RE-q}
\frac{\partial H}{\partial q} & =0.
\end{align}
\end{subequations}

Equations \eqref{eq:RE-p} and \eqref{eq:RE-m} are the kinematic equations (independent of the potential), while \eqref{eq:RE-q} is the dynamical equation.   We assume throughout that $V_\kappa'(q)\neq0$ for all $q$ (except in one case when $\kappa=0$ ---see the end of this section and Section\,\ref{S:attracting-repelling}).

 These equations were analysed in \cite{BGMM} for $\kappa\neq0$, but we outline the calculations again here, following
 a  different approach that is more convenient for our purposes. 
First fix $q\in I_\kappa$.  Combining equations\,\eqref{eq:RE-p} and \eqref{eq:RE-q} shows that 
$$m_1 = \frac{\mu_1+\mu_2}{\mu_2}\,p,\quad \text{and}\quad 
	p\bigl(\mu_2 \Sk(q)m_2-(\mu_1+\mu_2)\Ck(q)m_3\bigr)=0.$$
Hence either $p=m_1=0$ or there are linear relations between $m_2$ and $m_3$, and between $p$ and $m_1$.

If $p\neq0$ then the remaining equations lead to $V'(q)=0$, which we are assuming is not satisfied. 

Now assume $p=m_1=0$, which implies that~\eqref{eq:RE-p} and
 the second and third components of~\eqref{eq:RE-m} hold.  The dynamical equation~\eqref{eq:RE-q} is equivalent to  
 \begin{equation}
 \label{eq:m2asfunctionofm3}
m_2=\frac{(\mu_1+\mu_2)\Ck(q)m_3^2-\Sk^3(q)V_\kappa'(q)}{\mu_2 \Sk(q)m_3}.
\end{equation}
Using this, one may eliminate $m_2$ from the first component of~\eqref{eq:RE-m} and obtain a biquadratic equation for  $m_3$,  
 \begin{equation*}
am_3^4+bm_3^2+c=0
\end{equation*}
 whose coefficients $a,b$ and $c$ depend  on  $q, \kappa$ and the masses. Denote by $\mu$ the mass ratio $\mu=\mu_1/\mu_2$, and for  the rest of the paper we assume, without  any loss of generality, that $0<\mu \leq 1$  and $\mu_2=1$.

 The solution of this biquadratic equation leads to the following conditions for \RE:
 \begin{description}
\item[1.]  If either $\kappa<0$ or $\kappa>0$ and $q\neq\pi/(2\sqrt{\kappa})$ then $m_3^2=A_\pm(q, \kappa, \mu)$ where:
\begin{equation}
\label{eq:A}
A_\pm(q, \kappa, \mu)=-\frac{\left ( 2\mu \Ck^2(q)+1-\mu  \pm \sqrt{4\mu\Ck^2(q) + (\mu-1)^2} \right )\Sk(q) V_\kappa'(q)}{2\mu\kappa\Ck(q)}.
\end{equation}

\item[2.] If $\kappa > 0$ and $q={\pi}/{2\sqrt{\kappa}}$, then $a=b=0$  and we obtain the condition
\begin{equation}
\label{eq:right-angle}
(\mu-1)V_\kappa'(q)=0
\end{equation}
which is independent of $m_3$ and only has solutions if the masses are equal.

\item[3.] If $\kappa = 0$ then $a=0$, and we are left with a   linear equation
whose unique solution is 
\begin{equation}
\label{eq:m3flat}
m_3^2=\frac{\mu q^3}{1+\mu} V_0'(q).
\end{equation}
\end{description}

In case 3 above, the value of $m_3^2$ is the limiting value of $A_-$ as $\kappa\to 0$.   In case 1  we have $m_3^2=A_\pm(q, \kappa, \mu)$ so there are relative equilibria
provided that $A_\pm(q, \kappa, \mu)$ is positive.  The following lemma gives necessary and sufficient conditions for this.  Recall that $q\sqrt\kappa$ is a dimensionless quantity. 

\begin{lemma}
\label{L:A}
For any value of\/ $\mu\in(0,1]$, the quantity $ A_\pm(q, \kappa, \mu)$ defined by~\eqref{eq:A} has the following properties:
\begin{description}
\item Suppose the potential is attractive, $V_\kappa'(q)>0$, then:
\begin{description}
\item if $\kappa<0$  then $A_\pm(q, \kappa, \mu)$ is positive for all $q>0$;
\item if $\kappa >0$  then $A_-(q, \kappa, \mu)$ is positive if and only if $q<\pi/2\sqrt{\kappa}$, 
and $A_+(q, \kappa, \mu)$ is positive if and only if $\pi/2<q\sqrt{\kappa}<\pi$.
\end{description}
\item Suppose, on the other hand, the potential is repelling, $V_\kappa'(q)<0$, then
\begin{description}
\item if $\kappa<0$  then $A_\pm(q, \kappa, \mu)$ is negative for all $q>0$;
\item if $\kappa >0$  then $A_+(q, \kappa, \mu)$ is positive if and only if $q\sqrt{\kappa}<\pi/2$, 
and $A_-(q, \kappa, \mu)$ is positive if and only if $\pi/2<q\sqrt{\kappa}<\pi$.
\end{description}
\end{description}
\end{lemma}

The proof is postponed to the end of the section. We now state two  propositions classifying the relative equilibria for attracting and repelling potentials. 
 Our labelling of the \RE\  uses the terminology  of~\cite{BGMM}. In our classification
we do not distinguish \RE\  related by the time reversibility of the problem.

\begin{proposition}[Classification of relative equilibria for an attractive potential]
\label{P:RE-attractive}
In the presence of an attractive potential  $V_\kappa'(q)>0$,  the classification of 
relative equilibria of the problem 
is as indicated below. 
\begin{description}
\item If $\kappa<0$  then for any $q>0$ there are exactly two  relative equilibria respectively
having
\begin{equation*}
\begin{split}
m_3^2&=A_+(q, \kappa, \mu) \qquad \mbox{(\defn{hyperbolic \RE})}, \\
m_3^2&=A_-(q, \kappa, \mu)\qquad \mbox{(\defn{elliptic \RE}).}
\end{split}
\end{equation*}
For both relative equilibria $p=m_1=0$ and $m_2$ is determined by~\eqref{eq:m2asfunctionofm3}.

\item If $\kappa >0$  then for any $0<q<\pi/\sqrt{\kappa}$, $q\neq \pi/2\sqrt{\kappa}$, there
is exactly one relative equilibrium having
\begin{equation*}
\begin{split}
m_3^2&=A_+(q, \kappa, \mu) \quad \mbox{if $\quad \pi/2<q\sqrt{\kappa}<\pi$} 
\quad \mbox{(\defn{obtuse attractive \RE})}, \\
m_3^2&=A_-(q, \kappa, \mu) \quad \mbox{if $\quad 0< q\sqrt{\kappa}< \pi/2$} 
\quad \mbox{(\defn{acute attractive \RE})},
\end{split}
\end{equation*}
In  both cases $p=m_1=0$ and $m_2$ is determined by~\eqref{eq:m2asfunctionofm3}.

\item If $\kappa >0$  and $q \sqrt{\kappa} =\pi/2$, then a relative equilibrium is possible only 
if the two masses are equal. In this case there is a family of relative equilibria parametrized by
$m_3$ having $p=m_1=0$ and $m_2$ given by~\eqref{eq:m2asfunctionofm3}. These are 
\defn{right-angle
attractive \RE}.

\item If $\kappa =0$ then for any $q>0$ there is exactly one relative equilibrium determined
by
\begin{equation*}
m_3^2=\frac{\mu q^3}{1+\mu} V_0'(q), \qquad m_2=m_1=p=0.
\end{equation*}
These are the usual \defn{Keplerian \RE}: two bodies rotating uniformly about their centre of mass.
\end{description}
\end{proposition}
\begin{proof}
The proof follows from Equations~\eqref{eq:A}, \eqref{eq:right-angle},  \eqref{eq:m3flat} 
and Lemma~\ref{L:A}.
\end{proof}

The classification for a repelling potential (similar to the result of Section\,\ref{sec:repelling RE}) is given by the following.

\begin{proposition}[Classification of relative equilibria for a repelling potential]
\label{P:RE-repelling}
In the presence of a repelling potential  $V_\kappa'(q)<0$,  the classification of 
relative equilibria of the problem 
is as follows. 
\begin{description}
\item If $\kappa \leq 0$  there are no relative equilibria.

\item If $\kappa >0$  then for any $q$ satisfying $0<q<\pi/\sqrt{\kappa}$, $q\neq \pi/(2\sqrt{\kappa})$, there
is exactly one relative equilibrium having
\begin{equation*}
\begin{split}
m_3^2&=A_-(q, \kappa, \mu) \quad \mbox{if $\quad \pi/2<q\sqrt{\kappa}<\pi$} 
\quad \mbox{(\defn{obtuse repelling \RE})}, \\
m_3^2&=A_+(q, \kappa, \mu) \quad \mbox{if $\quad 0< q\sqrt{\kappa}< \pi/2$} 
\quad \mbox{(\defn{acute repelling \RE})},
\end{split}
\end{equation*}
In  both cases $p=m_1=0$ and $m_2$ is determined by~\eqref{eq:m2asfunctionofm3}.

\item If $\kappa >0$  and $\sqrt{\kappa}\,q =\pi/2$, then a relative equilibrium is possible only 
if the two masses are equal. In this case there is a family of relative equilibria parametrized by
$m_3$ having $p=m_1=0$ and $m_2$ given by~\eqref{eq:m2asfunctionofm3}. These are 
\defn{right-angled repelling \RE}.

\end{description}
\end{proposition}

Note that our classification  results in the repelling case for $\kappa>0$  may be recovered from the results
for the attractive case together with Lemma~\ref{lemma:anitpodal}.

We finish the section with a proof of Lemma~\ref{L:A}.

 \begin{proof} (of Lemma~\ref{L:A})
 We only treat the attractive case where $V_\kappa'(q)>0$, the other is analogous.
 
 It is clear that the numerator of $A_+(q, \kappa, \mu)$ is positive, so its sign is given by the
 denominator. For $\kappa<0$ we have $\Ck(q)>0$ for all $q>0$. On the other hand, 
  for $\kappa>0$,
 we have $\Ck(q)>0$ for $0<q\sqrt{\kappa}<\pi/2$ and $\Ck(q)<0$ for 
 $\pi/2<q\sqrt{\kappa}<\pi$, and the conclusions follow from these considerations.
 
 To analyse the sign of $A_-(q, \kappa, \mu)$ we start from the identity
 \begin{equation*}
\left ( 4\mu\Ck^2(q) + (\mu-1)^2 \right ) - \left ( 2\mu \Ck^2(q)+1-\mu \right )^2
=4\kappa \mu^2\Ck^2(q)\Sk^2(q).
\end{equation*}
From this equation it is straightforward to obtain 
\begin{equation*}
\mbox{sign} \left (  2\mu \Ck^2(q)+1-\mu  - \sqrt{4\mu\Ck^2(q) + (\mu-1)^2} \right ) = -\mbox{sign}(\kappa),
\end{equation*}
and the conclusions about the sign of $A_-(q, \kappa, \mu)$ follows 
 from the above considerations about the sign of $\Ck(q)$.
 \end{proof}

\begin{remark}\label{rmk:no potential}
Extending Proposition\,\ref{P:RE-flat} above, if, for arbitrary $\kappa$, and some particular value $q_0$ of $q$, the potential satisfies $V_\kappa'(q_0)=0$ then it follows from \eqref{eq:m2asfunctionofm3} that \RE\  occur with
$$p=\frac{m_1}{1+\mu},\qquad  m_2 = (\mu+1)\,\cot_\kappa(q_0)\,m_3.$$
Such motion corresponds to the two particles lying at a distance where the force vanishes, and following respective geodesics with equal speeds maintaining this constant separation.  For $\kappa=0$, this amounts to the particles having equal velocity.  
\end{remark}

\subsubsection{Flat case: $\kappa=0$} \label{sec:flat case Kepler}
Again it is useful to discuss how the the results above relate to the well-known properties of the 2-body problem in the plane.  First if the interaction is repelling, Proposition\,\ref{P:RE-repelling} tells us there are no relative equilibria, as is to be expected: indeed there are no motions where the particles remain at a constant distance. 

More interesting is the case of an attracting potential; see  Proposition\,\ref{P:RE-attractive}.  The Keplerian \re\ consist of uniform rotations about a fixed centre of mass.   However, notice that  there are motions where the particles remain at a constant distance that are not relative equilibria: namely, those motions where the particles rotate uniformly about the centre of mass, but the centre of mass is uniformly translating relative to the (fixed) plane. These become \re\ if one includes Galilean ssymmetry, but since such symmetries do not extend to the curved surfaces it is not helpful in this analysis. 

Finally, for the purposes of  section~\ref{S:attracting-repelling}, we consider the motion with vanishing potential. 
In the absence of any interaction, the two particles will perform independent rectilinear motion.  These are \re\ precisely when they have identical velocities; up to Euclidean symmetry, there is a two parameter family of such motions, classified in terms of the reduced variables in the following proposition, whose proof follows immediately from the analysis of~\eqref{eq:RE} with the simplifications
$\sin_0(q)=q$, $\cos_0(q)=1$ and $V_0'(q)=0$.  

\begin{proposition}[Classification of relative equilibria in the absence of interaction for $\kappa=0$]
\label{P:RE-flat}
In the absence of potential $V$, and for $\kappa=0$  and any  $q>0$ and $0< \mu \leq 1$ there is a two-parameter
family of \RE\  of the problem determined by the conditions
\begin{equation}
\label{eq:flat-RE2}
	p=\frac{m_1}{1+\mu},\qquad  m_2 = \frac{\mu+1}{ q}\,m_3.
\end{equation}
\end{proposition}

Following \eqref{eq:relations between variables}, the first relation signifies $\dot q=0$, while the second is equivalent to vanishing angular momentum about the centre of mass ($L=0$  in \eqref{eq:L=ang mmtm}). 
We are especially interested in the special class of \RE\  having $p=m_1=0$: those where the velocities are perpendicular to the line joining the particles as these turn out to be the \re\ arising as limits of \RE\  with $\kappa\neq0$, the so-called `perpendicular \re' in the introduction and in Section\,\ref{S:attracting-repelling} below.

\section{Attracting family} 
\label{S:attracting}
We consider here the family of \re\  that extend the 
 Keplerian \re\ in the plane. The 
  interaction potential $V_\kappa$ is assumed to be always attractive and to depend smoothly on $\kappa$. This is related to the work \cite{CRS} concerning the variation of the Kepler problem as the curvature is varied through 0.

\subsection{The family of relative equilibria}

Suppose $V_\kappa(q)$ is a smooth family of functions (i.e., smooth in both $q$ and $\kappa$) defined for $q\in I_\kappa$, and such that for each $\kappa$,  $V_\kappa$ is an increasing function of $q$.  The canonical example would be the graviational potential
$$V_\kappa(q) = -G\cot_{\kappa}(q)$$
where the constant $G>0$ encodes the gravitational constant and the masses.
Since $V_\kappa'(q)>0$, the  terminology and relevant conclusions of the previous section come
from Proposition \ref{P:RE-attractive}.

\begin{theorem} 
\label{Th:RE-transition-attractive}
Let $V_\kappa(q)$ be a smooth family of attractive potentials i.e. $V_\kappa'(q)>0$ for all
$q$ and $\kappa$.
Fix $0<\mu\leq 1$ and  $q>0$, and consider $\kappa \in (-\infty, (\pi/2q)^2)$. 
Then, for the given values of $q$ and $\mu$,  there
is a smooth transition from the  elliptic \RE\  
for $\kappa<0$ to  the 
acute attractive \RE\  for $\kappa>0$.  These \RE\  are interpolated at $\kappa=0$
by the 
Keplerian \RE\ .
\end{theorem}
 \begin{proof}
 The \RE\  under consideration have $m_3^2=A_-(q,\kappa,\mu)$ if $\kappa\neq 0$ and 
 $m_3^2=  \frac{\mu q^3}{1+\mu} V_0'(q)$ if $\kappa=0$. For all of them $m_2$ is
 given by~\eqref{eq:m2asfunctionofm3} and $p=m_1=0$.
 
 The smooth dependence of these \RE\  away from  $\kappa=0$ is clear from the expression
 for $A_-(q,\kappa,\mu)$ in~\eqref{eq:A}. In particular, for $\kappa\neq 0$, the denominator does not
 vanish by our assumption that $\kappa<(\pi/2q)^2$. The behaviour of $A_-(q,\kappa,\mu)$ in
 the vicinity of $\kappa=0$ is analysed using the series expansions~\eqref{eq:Taylor}.
 We obtain
 \begin{equation*}
A_-(q,\kappa,\mu)= \frac{\mu q^3}{1+\mu} V_\kappa'(q) - \frac{\mu q^5}{(1+\mu)^3} V_\kappa'(q) \kappa + \mathcal{O}(\kappa^2) \qquad \mbox{as} \quad  \kappa \to 0.
\end{equation*}
This proves that $A_-(q,\kappa,\mu)$ can be extended smoothly to $\kappa=0$ and moreover
that such extension satisfies
\begin{equation*}
A_-(q,0,\mu)= \frac{\mu q^3}{1+\mu} V_0'(q),
\end{equation*}
which is the value of $m_3^2$ for the Keplerian \RE\ . The proof is completed by noting that~\eqref{eq:m2asfunctionofm3} depends smoothly on $\kappa$.
 \end{proof}

\begin{figure}
\centering
\subfigure[$\kappa=0.2$   ]{\includegraphics[height=5.5cm]{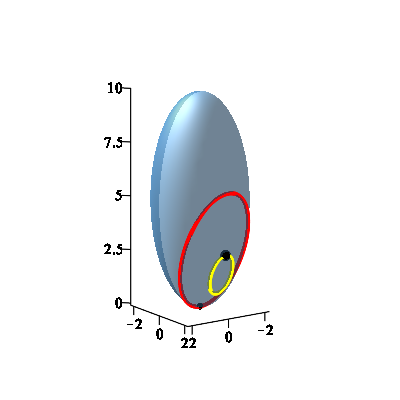}}  \hskip-5mm
\subfigure[$\kappa=0$ ]{\includegraphics[height=5cm]{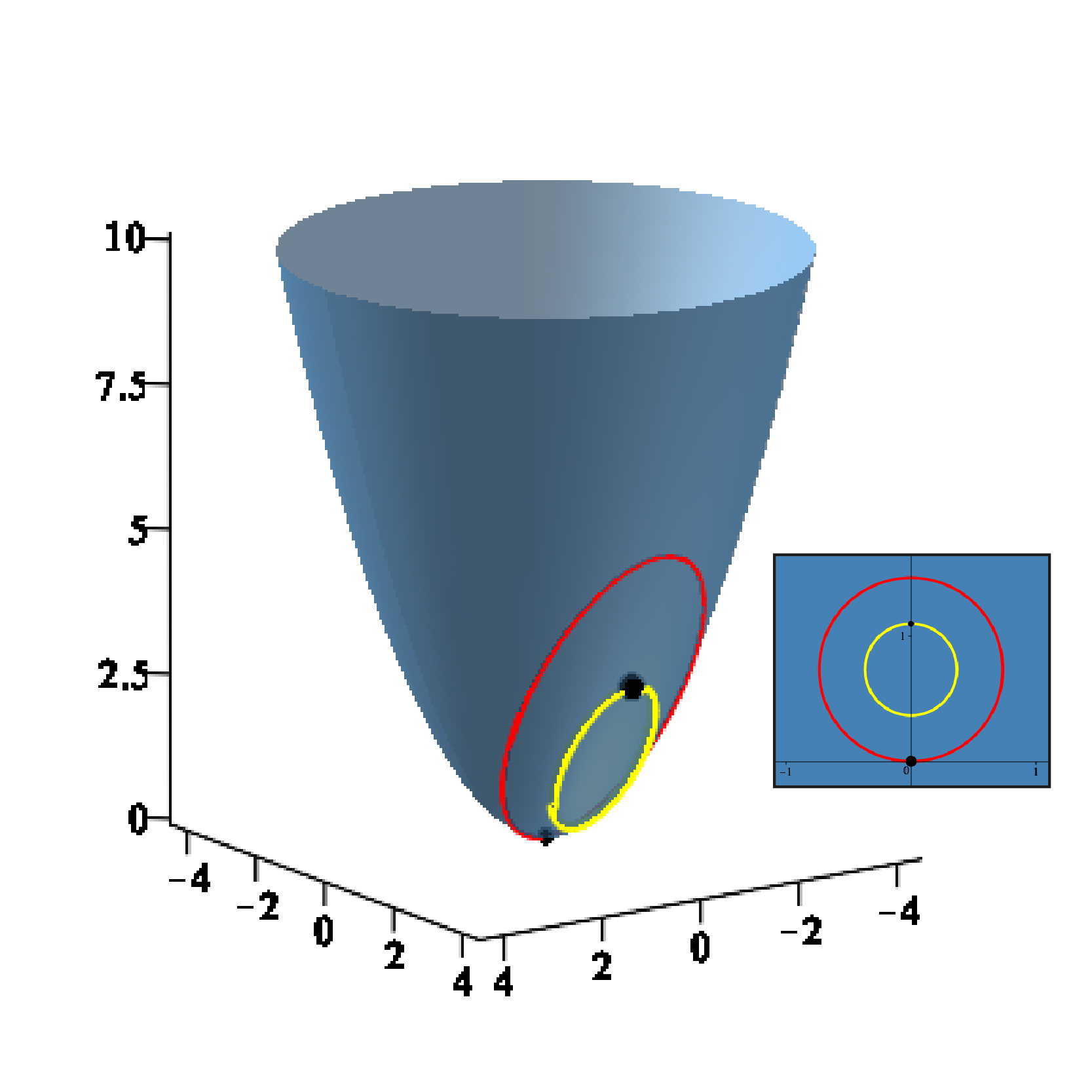}}  
\subfigure[$\kappa=-0.2$ ]{\includegraphics[height=5cm]{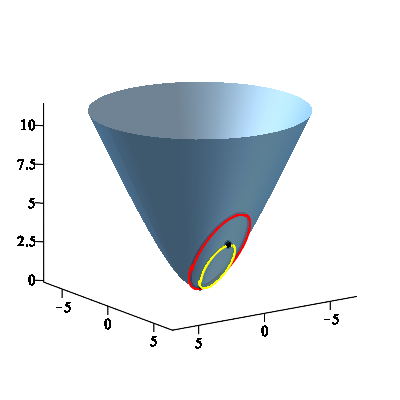}}
\caption{Family of \RE\  of Theorem~\ref{Th:RE-transition-attractive}
 for the potential $V_\kappa(q) = - \cot_\kappa(q)$ with
$\mu=0.5$ and $q=2.5$.  For $\kappa=0$, the inset figure shows the orthogonal projection of the trajectories onto the plane which are  concentric circles.  
} \label{fig:A-A family of RE}
\end{figure}

Figure\,\ref{fig:A-A family of RE} illustrates how the \RE\  in the family of the theorem 
 vary as the value  of the curvature passes from positive to negative.
  
 We finish the section by noting that the value of the Casimir function $C$ given by~\eqref{eq:Casimir} has the following asymptotic expansion 
 along the  \RE\  considered in this section:
 \begin{equation}
C=\frac{\mu q^3}{1+\mu}V_\kappa'(q)\kappa +\mathcal{O}(\kappa^2), \qquad \mbox{as} \quad \kappa \to 0.
\end{equation}
Therefore, in our convention, the sign of the Casimir coincides with the sign of the curvature along this family.

  \subsection{Linearization \& stability}
 
 Now consider the potential $V_\kappa(q)=-\cot_\kappa(q)$, and in particular $V_0$ is the Newtonian potential $-1/q$.
 
 The linearization of the reduced equations around the equilibrium having $q=q_0$, $p=m_1=0$,  and 
 \begin{equation*}
 \begin{split}
 m_3=\sqrt{A_-(q_0,\kappa,\mu)},   
\qquad m_2=\frac{\Ck(q_0)(\mu+1)A_-(q_0,\kappa,\mu)-\mu\Sk(q_0)}{\sqrt{A_-(q,\kappa,\mu)} \Sk(q_0)}, 
\end{split}
\end{equation*}
has the form
\begin{equation*}
\frac{\d}{\d t} {\bf w} = L(q_0,\mu, \kappa) {\bf w}, \qquad {\bf w}=(q,p,m_1,m_2,m_3)^t,
\end{equation*}
where the $5\times 5$ matrix $L(q_0,\mu, \kappa)$ has the following  asymptotic expansion as $\kappa\to 0$:
\begin{equation*}
L(q_0,\mu, \kappa)= L_0(q_0,\mu)+ L_1(q_0,\mu)\kappa +  \mathcal{O}(\kappa^2).
\end{equation*}
Here
\begin{equation*}
 L_0(q_0,\mu)=\begin{pmatrix}  0 & \frac{\mu+1}{\mu} & -\frac{1}{\mu} & 0 & 0 \\
\rule[-15pt]{0pt}{24pt}  -\frac{1}{q_0^3} & 0 & 0 & -\frac{1}{q_0^{3/2}\sqrt{\mu(\mu+1)} }& \frac{2\sqrt{\mu+1}}{q_0^{5/2}\sqrt{\mu}} \\
 0 & 0 & 0 & \frac{\sqrt{\mu+1}}{q_0^{3/2}\sqrt{\mu}} & 0 \\
\rule[-12pt]{0pt}{24pt}  0 & 0 & - \frac{\sqrt{\mu+1}}{q_0^{3/2}\sqrt{\mu}}  & 0 & 0 \\
 0 & 0 & -\frac{1}{q_0^{1/2}\sqrt{\mu(\mu+1)} } & 0 & 0
 \end{pmatrix},
\end{equation*}
and
\begin{equation*}
 L_1(q_0,\mu)=\begin{pmatrix}  0 &0 & 0 & 0 & 0 \\
\rule[-12pt]{0pt}{24pt}  \frac{-\mu(\mu+2)}{q_0(\mu+1)^2} & 0 & 0 & 
 	-\frac{\sqrt{\mu q_0} (\mu+2) }{2(\mu+1)^{5/2} }& 
	\frac{\sqrt{\mu+1}}{3\sqrt{\mu q_0}} \\
 \frac{1}{q_0(\mu+1)} & 0 & 0 &\frac{\sqrt{\mu q_0} (\mu - 2) }{2(\mu+1)^{3/2} }& \frac{1}{\sqrt{\mu(\mu+1)q_0}} \\
\rule[-12pt]{0pt}{24pt}  0 & -\sqrt{\frac{q_0}{\mu(\mu+1)}} & -\sqrt{\frac{q_0}{\mu}} \left ( \frac{\mu^2-2\mu-2}{2(\mu+1)^{3/2}} \right )  & 0 & 0 \\
 0 & -\frac{q_0^{3/2}}{(\mu+1)^{3/2}\sqrt{\mu}} & \frac{q_0^{3/2}(\mu^2+2\mu+4)}{6\sqrt{\mu} (\mu+1)^{5/2}} & 0 & 0
 \end{pmatrix}.
\end{equation*}
 
 The characteristic polynomial of $L(q_0,\mu,\kappa )$ defined by
 \begin{equation*}
p(x)=\det(L(q_0,\mu, \kappa) -x \mbox{Id}_5)
\end{equation*}
may be written as
\begin{equation*}
p(x)=-x \left ( x^2+ \frac{\mu+1}{\mu q_0^3}  + \mathcal{O}(\kappa^2)\right )  \left ( x^2+ \frac{\mu+1}{\mu q_0^3} + \frac{2\kappa (\mu^2+1)}{(\mu+1)\mu q_0} + \mathcal{O}(\kappa^2)
  \right ) .
\end{equation*}

\begin{figure}
\centering 
\includegraphics{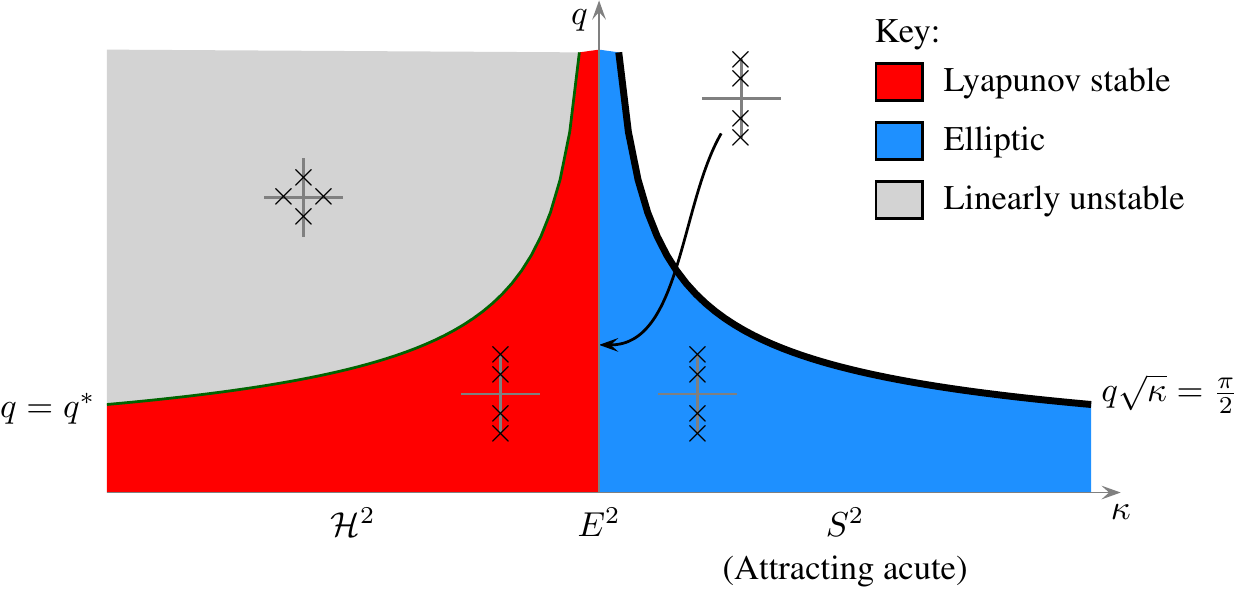}
\caption{Stability and eigenvalue pattern for \re\ of Kepler type in the attracting family with distinct masses, with $V_\kappa(q)= - \cot_\kappa(q)$.  See Remark\,\ref{rmk:q*} for the definition of $q^*$.} 
\label{fig:attracting-attracting}
\end{figure}

For small values of $\kappa$ one sees that all the four non-zero eigenvalues are imaginary (and double when $\kappa=0$:
 we  expect that to be a feature of the $1/r$ potential).  See Figure\,\ref{fig:attracting-attracting}.

The asymptotics of these eigenvalues as $\kappa\to 0$ are
$$ \pm i\left(\sqrt{\frac{\mu+1}{\mu q_0^3}} + \mathcal{O}(\kappa^2)\right), \quad
\text{and}\quad 
 \pm i\left(\sqrt{\frac{\mu+1}{\mu q_0^3}} + \sqrt{\frac{q_0}{(\mu+1)^3\mu }}(1+\mu^2)\kappa + \mathcal{O}(\kappa^2)\right).
$$
The double eigenvalues therefore separate with bounded speed in $\kappa$, as $\kappa$ moves from 0.

We end this section by addressing the question of why the eigenvalues at $\kappa=0$ are double, by relating the motion to the known behaviour of the Kepler problem.  
Firstly, for $\kappa=0$, the \re\ in question satisfies $p=m_1=m_2=0$, so is at a singular point of the Casimir, which for $\kappa=0$ is $C=m_1^2+m_2^2=\|\mathbf{p}\|^2$, see Sec.\,\ref{sec:flat case momentum}, and satisfies
$$m_3^2 = \frac{\mu q}{1+\mu}.$$
Let $\widehat\xx = (\widehat q,\widehat p,\widehat m_1,\widehat m_2,\widehat m_3)$  be a tangent vector at the (relative) equilibrium in the Poisson reduced space $\R^5$. The kernel of $L_0(q_0,\mu)$ is spanned by the vector $\widehat\xx=(2\sqrt{q_0(\mu+1)},\,0,\,0,\,0,\,\sqrt{\mu})^t$, which is tangent to the curve of \re.    For the non-zero eigenvalues, consider first the effect of varying $(m_1,m_2)$ away from 0. In the original dynamics (on $\R^4$), this amounts to increasing the momentum $\mathbf{p}$. The resulting motion is a superposition of the Kepler rotation and a uniform translation of the centre of mass.  In the reduced space, this is a periodic orbit, with period equal to the period of rotation of the Kepler solution.  

On the other hand, if we remain on the subset $m_1=m_2=0$ (where the Casimir vanishes, which is therefore an invariant submanifold), the function $m_3$ coincides with the angular momentum $L$ and so is conserved. Putting $m_3=\ell$ (a constant) we have a reduced 1 degree of freedom system in $(q,p)$ which coincides with the usual reduction with amended potential,
$$V_{\mathrm{am}}(q) = V_0(q) + \frac{A}{q^2},$$
for some constant $A$ depending on $\mu$ and $\ell$. The small oscillations near the equilibrium of this reduced system consists of elliptical orbits in the original system, and these are periodic with the same period as the circular motion.  

Thus all the periodic motions have the same period as the period of the circular Keplerian orbit represented by the \re\ in question.

\begin{remark}\label{rmk:q*}
While our primary interest in this paper involves the bifurcations arising as $\kappa$ passes through 0,  Figures\,\ref{fig:attracting-attracting} and \ref{fig:attracting-repelling} show some other stability changes. These are specifically for the potential $V_\kappa(q)=-\cot_\kappa(q)$ and $V_\kappa(q)=\kappa\cot_\kappa(q)$ respectively, and follow respectively from \cite{BGMM} and Theorem~\ref{Th:DistinctMasses}, adapted using the rescaling of Remark\,\ref{rmk:rescaling curvature}. The transition between (linearly) stable and unstable \re\ occurs respectively when $q=q^*$ and $q=q^\dagger$.  For the potential $V_\kappa(q)=-\cot_\kappa(q)$ treated in this section, the transition occurs when $\kappa<0$ and $q^*$ satisfies
\begin{equation}\label{eq:q*}
	q^* = \alpha^* + \frac{1}{2}\sin_\kappa^{-1} (\mu \sin_\kappa 2\alpha^*),
\end{equation}
where $\alpha^*$ is the unique solution to  the equation
\begin{equation}\label{eq:alpha*}
\cos_\kappa2\alpha=2|\kappa|\sin_\kappa^2\alpha \sqrt{1-\kappa\mu^2\sin_\kappa^2 2\alpha }.
\end{equation}
On the other hand, in Section\,\ref{S:attracting-repelling} below we consider  $V_\kappa(q) = \kappa\cot_\kappa(q)$, and then the transition occurs when $\kappa>0$ and the interaction is repelling. In this case, the bifurcation will occur for the  acute configuration at $q^\dagger$ 
given by the appropriate modification of \eqref{eq:q-dagger},
$$	q^\dagger=\alpha^\dagger -\frac\pi{2\sqrt{\kappa}} - 
			\frac{1}{2}\sin_\kappa^{-1} (\mu\sin_\kappa 2\alpha^\dagger)
$$
where $\alpha^\dagger\in[\pi/2\sqrt{\kappa},\,\pi/\sqrt{\kappa}]$ satisfies \eqref{eq:alpha*}.   
\end{remark}

\begin{remark}\label{rmk:nonlinear stability} Figure \ref{fig:attracting-attracting} shows that, for $\kappa<0$, the \re\ of elliptic type are Lyapunov stable, but on the other hand for $\kappa>0$ they are merely elliptic.  However, calculations in \cite[Sec.\,4.2]{BGMM} involving a KAM argument, show that for small values of $q\sqrt{\kappa}$ these are also Lyapunov stable (see also Fig.\,11 there).  For the intermediate value $\kappa=0$, the circular Kepler  orbits are well-known to be Lyapunov stable relative to $\SE(2)$. 
\end{remark}

\section{Attracting-repelling family} 
\label{S:attracting-repelling}

In this final section we consider the  family of  \re\  that extend the  `perpendicular \re' defined in the introduction. We will
assume that the interaction potential $V_\kappa(q)$ is 
repelling for $\kappa>0$, vanishes for $\kappa=0$ and is attracting for $\kappa<0$. Again, $V_\kappa$ should be defined for $q\in I_\kappa$. An example would be $V_\kappa(q) = \kappa \cot_{\kappa}(q)$.

\subsection{The family of relative equilibria}

The  terminology and relevant conclusions for our analysis of the  attracting-repelling family  come
from Propositions \ref{P:RE-attractive}, \ref{P:RE-repelling} and \ref{P:RE-flat}. The following
theorem shows that our assumption on the potential allows
us to connect smoothly the hyperbolic \RE\  for $\kappa<0$ with the acute repelling \RE\  for $\kappa>0$.

\begin{theorem} 
\label{Th:RE-transition-attractive-repelling}
Let $V_\kappa(q) = -\kappa U_\kappa(q) $ be a smooth family of potentials having 
$U_\kappa'(q)>0$ for all
$q$ and $\kappa$, so that  the interaction between the particles is repelling for $\kappa>0$, vanishes for 
$\kappa=0$ and is attracting for $\kappa<0$.

Fix $0<\mu\leq 1$ and  $q>0$, and consider $\kappa \in (-\infty, (\pi/2q)^2)$.
Then, for the given values of $q$ and $\mu$,  there is
a smooth transition from the  hyperbolic \RE\  
for $\kappa<0$ to the 
acute repelling \RE\  for $\kappa>0$. These \RE\  are interpolated at $\kappa=0$
by the \RE\  of the planar 2-body problem with no interaction satisfying
\begin{equation*}
m_3^2=\frac{(\mu+1)q}{\mu}\,U'_0(q), \qquad  m_2 = \frac{\mu+1}{ q}\,m_3, \qquad p=m_1=0.
\end{equation*}
\end{theorem}

With no interaction, all solutions of the planar 2-body problem consist of the 2 particles in rectilinear motion. These are \re\ (relative to the Euclidean group) precisely when their velocities are equal.  In addition, it follows from \eqref{eq:relations between variables} that the \re\ satisfy the conditions in the theorem if their common velocity is perpendicular to the line joining the particles (called `perpendicular motion' in the introduction).

 \begin{proof}
 For $\kappa \neq 0$, the \RE\  under consideration have $m_3^2=A_+(q,\kappa,\mu)$, $m_2$ 
 given by~\eqref{eq:m2asfunctionofm3} and $p=m_1=0$.
The expression~\eqref{eq:m2asfunctionofm3} may be evaluated at $\kappa=0$ and, under our 
assumption that $V_0'(q)=0$, it simplifies to $m_2 = \frac{\mu+1}{ q}\,m_3$. Hence, to complete the 
proof we  only need to show that $A_+(q,\kappa,\mu)$ admits a smooth extension at $\kappa=0$ 
and that this extension satisfies $A_+(q,0,\mu)=\frac{(\mu+1)q}{\mu}U'_0(q)$.
But this is obvious from~\eqref{eq:A} since our assumption on the form of $V_\kappa(q)$ cancels
the factor of $\kappa$ in the denominator of $A_+(q,\kappa,\mu)$ and  $A_+(q,0,\mu)$ may
be directly evaluated to obtain the given value.
 \end{proof}

\begin{figure}
\centering
\subfigure[$\kappa=0.2$  (repelling) ]{\includegraphics[height=5.5cm]{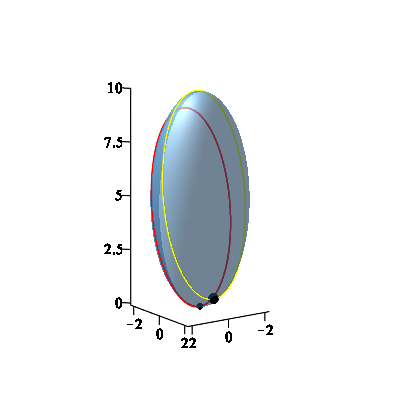}} \hskip-5mm
\subfigure[$\kappa=0$  (no interaction)]{\includegraphics[height=5cm]{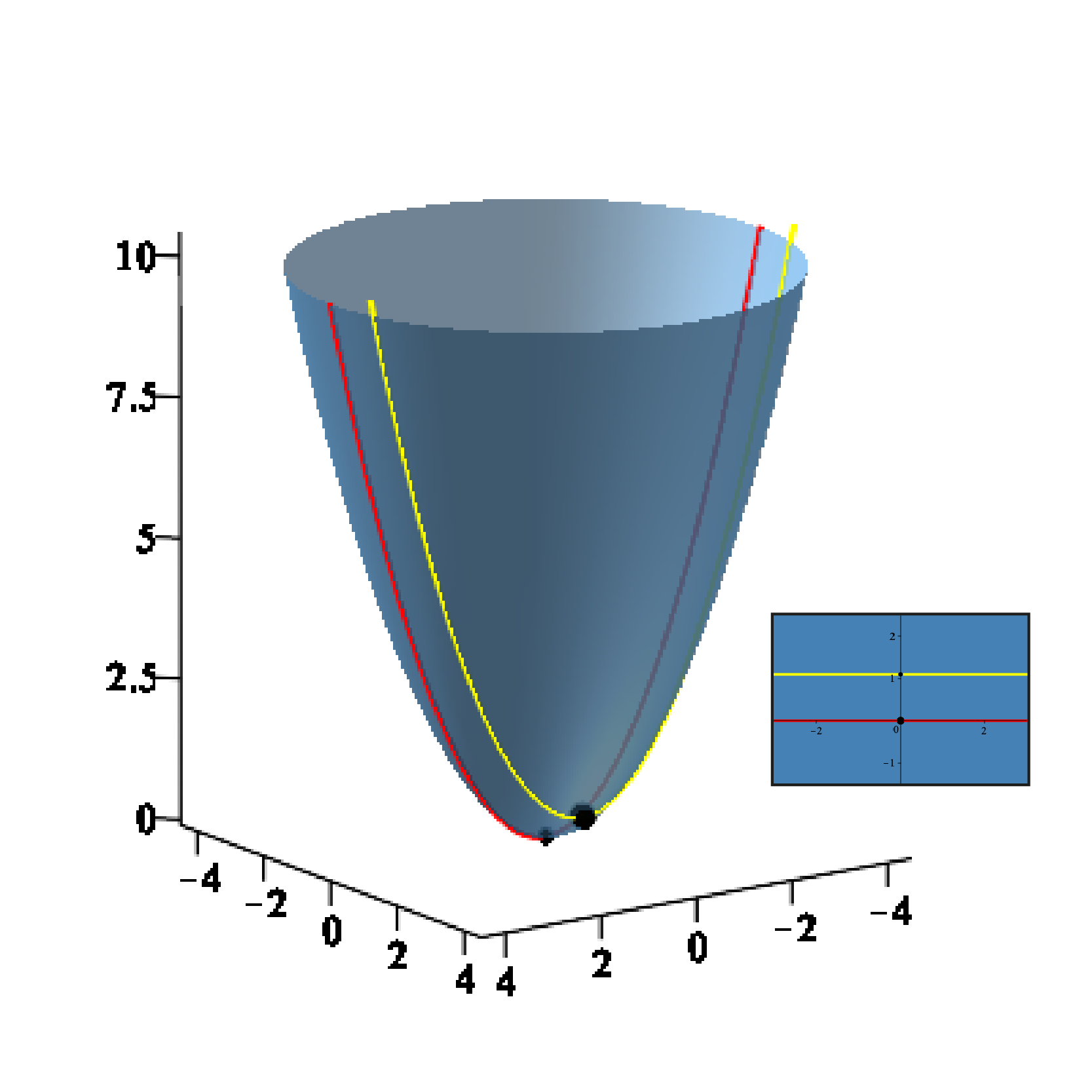}} 
\subfigure[$\kappa=-0.2$  (attracting)]{\includegraphics[height=5cm]{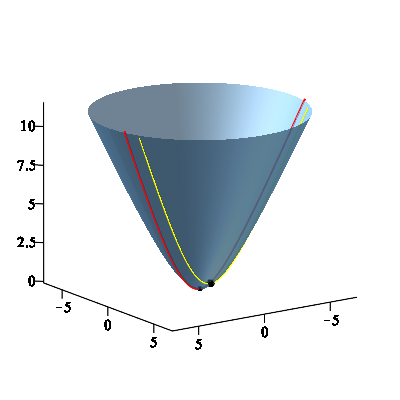}}
\caption{Family of \RE\  of Theorem~\ref{Th:RE-transition-attractive-repelling}
 for the potential $V_\kappa(q) = \kappa \cot_\kappa(q)$ with
$\mu=0.5$ and $q=1.1$.  In the case of $\kappa=0$, the inset figure shows the orthogonal projection of the two curves to the plane which are parallel lines.} \label{fig:A-R family}
\end{figure}

Figure\,\ref{fig:A-R family} illustrates how the \RE\  of the family described in the theorem 
 vary as  the value of the curvature passes from positive to negative.
  
We end up by giving the asymptotic expansion around $\kappa=0$ for the Casimir function
$C$ given by \eqref{eq:Casimir} 
along the family of \RE\  considered  in this section. Under the assumption that $V_\kappa=-\kappa U_\kappa(q)$
as in the statement of Theorem~\ref{Th:RE-transition-attractive-repelling} we have
\begin{equation*}
C= \left ( \frac{(\mu+1)^2U_\kappa'(q)}{\mu m_3}  \right )^2 + \mathcal{O}(\kappa) \qquad \mbox{as}
\qquad \kappa \to 0. 
\end{equation*}
Therefore the Casimir is positive for all members of this family.
 
 \begin{remark}
The theorem and proof are  easily generalised to allow more general potentials transitioning the
 attractive interaction for $\kappa<0$ with the repelling interaction for $\kappa>0$. In particular,
one may allow  $V'_\kappa(q)$ to vanish
at higher order in $\kappa$ at $\kappa=0$ in which case  one finds that the interpolating \RE\  at $\kappa=0$ have 
 $m_3=m_2=0$.
Namely, in such case,  the limiting \RE\  at $\kappa=0$ become equilibria where
the masses do not move  (and $C=0$ when $\kappa=0$).
\end{remark}
 
  \subsection{Linearization \& stability}
 
 Assume $V_\kappa(q)= \kappa \cot_\kappa(q)$.
 The linearization of the reduced equations around the equilibrium  at $q=q_0$, $p=m_1=0$ and
 \begin{equation*}
\quad m_3=\sqrt{A_+(q_0,\kappa,\mu)}\, , \quad m_2=\frac{\Ck(q_0)(\mu+1)A_+(q_0,\kappa,\mu)+
 \kappa \mu\Sk(q_0)}{\sqrt{A_+(q_0,\kappa,\mu)} \Sk(q_0)}, 
\end{equation*}
has the form
\begin{equation*}
\frac{\d}{\d t} {\bf w} = L(q_0,\mu,\kappa) {\bf w}, \qquad {\bf w}=(q,p,m_1,m_2,m_3)^t,
\end{equation*}
where the $5\times 5$ matrix $L(q_0,\mu,\kappa)$ has the following  asymptotic expansion as $\kappa\to 0$:
\begin{equation*}
L(q_0,\mu,\kappa)= L_0(q_0,\mu)+ L_1(q_0,\mu)\kappa + L_2(q_0,\mu)\kappa^2  +  \mathcal{O}(\kappa^3).
\end{equation*}
Here
\begin{equation*}
 L_0(q_0,\mu)=\begin{pmatrix}  0 & \frac{\mu+1}{\mu} & -\frac{1}{\mu} & 0 & 0 \\
\rule[-14pt]{0pt}{24pt} -\frac{(\mu+1)^2}{\mu^2q_0^5} & 0 & 0 & - \frac{ \sqrt{\mu+1}}{\mu^{3/2} q_0^{5/2}} &  \frac{ (\mu+1)^{3/2}}{\mu^{3/2} q_0^{7/2}} \\
\rule[-12pt]{0pt}{24pt}   -\frac{(\mu+1)^3}{\mu^2q_0^5}  & 0 & 0 & - \frac{ (\mu+1)^{3/2}}{\mu^{3/2} q_0^{5/2}}  &  \frac{ (\mu+1)^{5/2}}{\mu^{3/2} q_0^{7/2}}  \\
 0 & 0 &0  & 0 & 0 \\
 0 & \left ( \frac{\mu+1}{\mu q_0} \right )^{3/2} & - \frac{ \sqrt{\mu+1}}{(\mu q_0)^{3/2}} & 0 & 0
 \end{pmatrix}.
 \end{equation*}
Explicit expressions for $L_1$ and $L_2$ can be found by lengthy computations.  
 
 The characteristic polynomial of $L(q_0,\mu,\kappa)$ defined by
 \begin{equation*}
p(x)=\det(L(q_0,\mu,\kappa) -x \mbox{Id}_5)
\end{equation*}
may be written as
\begin{equation*}
p(x)=-x \left ( x^4+ b(q_0,\mu,\kappa )x^2 + c(q_0,\mu ,\kappa)  \right ).
\end{equation*}
with
\begin{equation*}
 b(q_0,\mu,\kappa)=\frac{2(\mu+1)}{\mu q_0^3} \kappa 
 + \mathcal{O}(\kappa^2), \qquad  
 c(q_0,\mu,\kappa)= -\frac{3(\mu+1)^2 \kappa^2}{\mu^2 q_0^6}+ \mathcal{O}(\kappa^3).
\end{equation*}

Note that $x=0$ is an eigenvalue of $L$ for all $\kappa$; this is due to the Casimir being an integral of motion. Indeed $L^t\,\nabla C=0$.

\begin{figure}
\centering 
\includegraphics{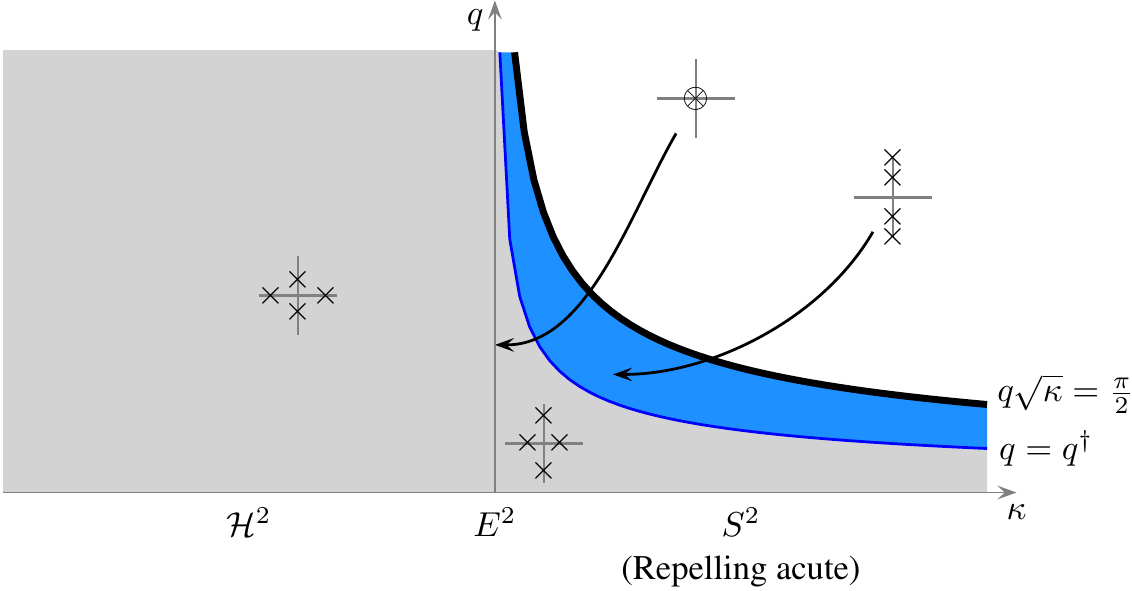}
\caption{Stability and eigenvalue pattern for \re\ of translation type in the attracting-repelling family with $V_\kappa(q)=\kappa \cot_\kappa(q)$. (See Fig.\,\ref{fig:attracting-attracting} for the colour key and Remark\,\ref{rmk:q*} for the definition of $q^\dagger$.)}
\label{fig:attracting-repelling}
\end{figure}

At $\kappa=0$, the all eigenvalues of the matrix $L_0$ vanish; indeed $L_0$ is nilpotent of rank 2 and index 2 ($L_0^2=0$---see below for an explanation).  As $\kappa$ moves away from zero, and for fixed $q=q_0$, one finds asymptotic expansions as $\kappa\to0$ for the 4 non-zero eigenvalues to be
\begin{equation*}
 \pm B_0\sqrt{\kappa} + \mathcal{O}(\kappa^{3/2})
 	 \quad\text{and}\quad
 \pm B_0\sqrt{-3\kappa} + \mathcal{O}(\kappa^{3/2}),
\end{equation*}
where $B_0^2=\frac{2(\mu+1)}{\mu q_0^3}$.  We see that, regardless of the sign of $\kappa$, two of these are real and two are imaginary as illustrated in Figure\,\ref{fig:attracting-repelling}.  They move away from 0 with unbounded speed, as one sees in traditional saddle-node or pitchfork bifurcations.

We describe briefly the geometry  of the nilpotency of $L_0$.  Fix $\kappa=0$, and consider the \re\ that is perpendicular motion. The set of all equilibria of the reduced equations is 3-dimensional, parametrized for example by $(q,p,m_3)$ and satisfying\,\eqref{eq:flat-RE2}. The tangent space to this subspace accounts for the 3-dimensional kernel of $L_0$.  Since $L_0$ is nilpotent of index 2, the image of $L_0$ lies in this subspace, and in particular coincides with the tangent space to the set of equilibria contained in the level set of the Casimir.

\subsection*{Acknowledgements}  The authors acknowledge support of  a Newton Advanced Fellowship from the Royal Society, Ref: NA140017 that financed the early stages of this research. LGN is thankful to the 
Alexander von Humboldt Foundation for a Georg Forster Advanced Research Fellowship   that funded 
a research visit to TU Berlin where part of his contribution to this research was done, 
and also to Department of Mathematics of the
 University of Manchester for its hospitality during his visit in April 2019.


\setlength{\parindent}{0pt}\small

\hrulefill

\bigskip

{L.C.~Garc\'ia-Naranjo} \\
Departamento de Matem\'aticas y Mec\'anica IIMAS-UNAM \\
Apdo. Postal: 20-726 Mexico City, 01000, Mexico\\
\texttt{luis@mym.iimas.unam.mx}

\bigskip

{J.~Montaldi} \\
Department of Mathematics, University of Manchester \\
Manchester M13 9PL, UK\\
\texttt{j.montaldi@manchester.ac.uk}

\end{document}